\documentclass{aa}  
\usepackage{longtable}
\usepackage{caption}
\usepackage{subcaption}
\usepackage{amsmath}
\usepackage{graphicx}
\usepackage{txfonts}
\usepackage[fleqn]{amsmath}	
\usepackage{amssymb}	
\usepackage{mathtools}
\usepackage[breaklinks=true]{hyperref}
\hypersetup{
  colorlinks=true,   
  citecolor=blue,    
  urlcolor=blue,     
  linkcolor=blue,     
}
\usepackage{lscape}
\usepackage{xcolor}

\newcommand{\lbol}{$\rm L_{bol}$}
\newcommand{\mbh}{$\rm M_{BH}$}

\newcommand{\Msun}{$\rm M_{\odot}$\xspace}
\newcommand{\Hb}{\ion{H}{$\beta$}\xspace}
\newcommand{\oiii}{[\ion{O}{iii}]\xspace}
\newcommand{\Feii}{\ion{Fe}{ii}\xspace}
\newcommand{\FeHb}{\ion{Fe}{ii}/\ion{H}{$\beta$}\xspace}
\newcommand{\FeMg}{\ion{Fe}{ii}/\ion{Mg}{ii}\xspace}

\DeclareRobustCommand{\ion}[2]{%
\relax\ifmmode
\ifx\testbx\f@series
{\mathbf{#1\,\mathsc{#2}}}\else
{\mathrm{#1\,\mathsc{#2}}}\fi
\else\textup{#1\,{\mdseries\textsc{#2}}}%
\fi}

\newcommand{\xmm}{XMM--\emph{Newton}\xspace}

\newcommand{\chandra}{\emph{Chandra}\xspace}

\newcommand{\jwst}{\textit{JWST}\xspace}

\newcommand{\xspec}{\textsc{xspec}\xspace}

\begin{document} 

   \title{GA-NIFS: an extended [OIII]$\lambda$5007 halo around the sub-Eddington quasar J1342+0928 at z=7.54}

  \author{Bartolomeo Trefoloni\inst{1,2}
  \thanks{\email{bartolomeo.trefoloni@sns.it}},
  Stefano~Carniani\inst{1},
  Elena~Bertola\inst{2},
  Giacomo~Venturi\inst{1},  
  Sandra~Zamora\inst{1},
  Eleonora~Parlanti\inst{1},
  Santiago~Arribas\inst{3}, 
  Andrew~Bunker\inst{4},
  St\'{e}phane~Charlot\inst{5},
  Francesco~D'Eugenio\inst{6,7}
  Peter~Jakobsen\inst{8,9},
  Roberto~Maiolino\inst{6,7,10},
  Michele~Perna\inst{3},
  Bruno~Rodr\'{i}guez~Del~Pino\inst{3},  
  Hannah~\"Ubler\inst{11},
  Chris~J.~Willott\inst{12},
  Torsten~B\"{o}ker\inst{13},
  Giovanni~Cresci\inst{2},
  Isabella~Lamperti\inst{14,2},
  Madeline~Marshall\inst{15},
  Pablo~G.~P\'{e}rez-Gonz\'{a}lez\inst{3}
  }
          
\institute{
 $^{1}$Scuola Normale Superiore, Piazza dei Cavalieri 7, I-56126 Pisa, Italy\\
 $^{2}$INAF -- Osservatorio Astrofisico di Arcetri, Largo Enrico Fermi 5, I-50125 Firenze, Italy\\
 $^{3}$Centro de Astrobiología (CAB), CSIC–INTA, Cra. de Ajalvir Km. 4, 28850 – Torrejón de Ardoz, Madrid, Spain\\
 $^{4}$Department of Physics, University of Oxford, Denys Wilkinson Building, Keble Road, Oxford OX1 3RH, UK\\
 $^{5}$Sorbonne Universit\'e, CNRS, UMR 7095, Institut d'Astrophysique de Paris, 98 bis bd Arago, 75014 Paris, France\\
 $^{6}$Kavli Institute for Cosmology, University of Cambridge, Madingley Road, Cambridge CB3 0HA, UK\\
 $^{7}$Cavendish Laboratory, University of Cambridge, 19 JJ Thomson Avenue, Cambridge CB3 0HE, UK\\
 $^{8}$Cosmic Dawn Center (DAWN), Copenhagen, Denmark.\\
 $^{9}$Niels Bohr Institute, University of Copenhagen, Jagtvej 128, DK-2200, Copenhagen, Denmark.\\
 $^{10}$ Department of Physics and Astronomy, University College London, Gower Street, London WC1E 6BT, UK\\
 $^{11}$Max-Planck-Institut f\"ur extraterrestrische Physik (MPE), Gie{\ss}enbachstra{\ss}e 1, 85748 Garching, Germany\\
 $^{12}$NRC Herzberg, 5071 West Saanich Rd, Victoria, BC V9E 2E7, Canada\\
 $^{13}$European Space Agency, c/o STScI, 3700 San Martin Drive, Baltimore, MD 21218, USA\\
 $^{14}$Dipartimento di Fisica e Astronomia, Universit\`a di Firenze, Via G. Sansone 1, 50019, Sesto F.no (Firenze), Italy\\
 $^{15}$Los Alamos National Laboratory, Los Alamos, NM 87545, USA\\
 \\}

\titlerunning{GA-NIFS: an extended [OIII] halo around the sub-Eddington quasar J1342+0928 at z=7.54}

\authorrunning{B. Trefoloni et al.}

\abstract{The James Webb Space Telescope (\textit{JWST}) opened a new observational window on the primordial Universe. Here we present new JWST NIRSpec integral field spectroscopy (IFS) observations of the $z=7.54$ quasar ULAS J1342+0928 obtained as part of the Galaxy Assembly with NIRSpec IFS (GA-NIFS) GTO programme. The new data-set obtained with both the prism ($R\sim100$) and the high-resolution grating ($R\sim2700$) allow for a complete description of the quasar emission from the rest-frame UV to optical bands. The low-resolution data reveal the presence of [\ion{O}{iii}] emission on $\sim$7 kpc scales, well above the typical galaxy size at this redshift, likely associated with a past outflow event. Additionally, the high-resolution observations show a more energetic ionised outflow on nuclear scales ($\lesssim 0.6$ kpc). The total ionised mass outflow rate ranges between 50 and 300 $\rm M_{\odot} \, yr^{-1}$ where the significant spread is mostly due to the lack of tight constraints on the electron density. This range overlaps in part with the star formation rate range (85--545 $\rm M_{\odot} \, yr^{-1}$), implying that the nuclear outflow could ultimately lead to an early star formation quenching. By employing an accretion disc modelling, for the first time on \textit{JWST} data, we manage to robustly estimate the black hole mass and the bolometric luminosity, $\rm \log(M_{BH}/(M_{\odot}))=9.2\pm 0.2$ and $\rm \log(L_{bol}/(erg \, s^{-1}))=46.8\pm 0.1$, respectively. We derive an Eddington ratio of $\rm \lambda_{Edd}\sim 0.4$, challenging the paradigm of widespread super-Eddington accretion in quasars at the epoch of reionisation. Lastly, we measure the most distant broad line ratios (\ion{Fe}{ii}\textsubscript{UV}/\ion{Mg}{ii} and \ion{Fe}{ii}\textsubscript{opt}/\ion{H}{$\beta$}) so far. Interpreting these ratios as proxies to the broad line region metallicity, we find evidence for an early metal enrichment in quasars within 700 Myrs from the Big Bang.}


   \maketitle
%

\section{Introduction}

Quasars (QSOs) are the most luminous persistent sources in the Universe, efficiently converting matter accreting onto supermassive black holes (SMBH) into light. The enormous amount of energy emitted by these powerful sources makes them stand as clear signposts of galaxy assembly at all cosmic times.

Several surveys operating during the last decade revealed (e.g. \citealt{wang2016survey, jiang2016final, matsuoka2019subaru, yang2023desi}) or further characterised (e.g. \citealt{farina2022x, yang2023spectroscopic}) high-redshift QSOs ($z\gtrsim5$), already active when the Universe was less than 1 Gyr old. More than 500 quasars have been discovered above $z=5$ so far (see also \citealt{fan2023quasars} for a recent review). Black hole masses ($\rm M_{BH}$) derived for these objects, are often found to exceed $\rm 10^9 \, M_{\odot}$ (e.g. \citealt{mazzucchelli2017physical, yang2023spectroscopic}).

However, the very existence of such massive objects in the epoch of reionisation poses serious challenges to current cosmological and accretion models. From a theoretical standpoint, it is hard to predict the size of the collapsed progenitors (seeds) and the route to accrete such a large mass (see e.g. \citealt{inayoshi2020assembly} for a review). In particular, models trying to explain how the aggregation of such masses might happen generally fall into two main categories, each with their own shortcomings with respect to observational constraints. A nearly continuous---unconventionally high---accretion on standard seeds, left by the first generation of stars, has been invoked to produce a $\rm 10^9 \, M_{\odot}$ SMBH within 1 Gyr (the `light seeds' scenario, e.g. \citealt{tanaka2009assembly, volonteri2010formation}). On the other hand, lower accretion rates could be enough to produce the observed masses in the case of initial `heavy seeds' ($\rm 10^4-10^6 \, M_{\odot}$, see e.g. \citealt{lodato2007mass}). In this picture, there must be one or more mechanisms that rapidly produce a $\rm 10^4 -10^6 \, M_{\odot}$ seed black hole, which then proceeds to accrete near the Eddington limit. Examples of such mechanisms are hyper-Eddington accretion onto a lower mass BH leading to a $\rm 10^5-10^6 \, M_{\odot}$ seed (\citealt{inayoshi2016hyper, ryu2016intermediate}), runaway collisions between stellar-mass BH and/or stars in dense proto-clusters (see e.g. \citealt{vergara2023global} and references therein), and the so-called direct-collapse black holes (DCBHs; e.g. \citealt{ferrara2014initial, inayoshi2014formation, sugimura2014critical}). Furthermore, primordial black holes (PBHs; \citealt{carr1974black, bean2002could, dolgov2018massive}) are another possibility to explain the large SMBH masses at high redshift, which has lately been gaining popularity (e.g. \citealt{suh2025super}). In this framework, PBHs would have formed during the radiation dominated era as a consequence of the gravitational collapse of overdense regions. Since PBHs originate at earlier cosmic times, they would have naturally had more time to accrete to the observed masses (\citealt{bernal2018signatures}).
In this context, reliably measuring SMBH masses at high redshift is a key requirement to place tighter constraints on the seeding mechanism.

The most robust technique to measure the black hole mass hosted by AGN is via reverberation mapping campaigns (e.g. \citealt{blandford1982reverberation, peterson1993reverberation, db2020, shen2024sloan}) generally targeting the H$\beta$ emission line in local sources. However, this technique becomes increasingly difficult at $z\gtrsim$1, with a few notable exceptions (e.g. \citealt{shen2014sloan, king2015simulations, saturni2016multi}) since the the H$\beta$ shifts into the infrared, as well as for the effect of cosmological time dilation, requiring increasingly longer monitoring campaigns. Furthermore, a smaller flux variability is observed in increasingly brighter sources such as quasars emitting at high redshift (e.g. \citealt{shen2024sloan}). Alternatively, a direct measurement of the differential phase of some broad emission lines from high-resolution infrared interferometry, combined with the modelling of the BLR dynamics, yielded an accurate estimate of the black hole mass even up to $z\sim2$ (e.g. \citealt{abuter2024dynamical}, but see also \citealt{gravity2018spatially} and \citealt{amorim2020spatially} for more local examples). Also this approach, due to technical reasons, has only been applied to a handful of suitable sources.

Because of these observational limits, the most widely employed technique for estimating the black hole mass in distant QSOs is via the single-epoch calibrations (SE; see e.g. \citealt{vestergaard2006determining} and references therein). This technique takes advantage of the tight relation between the broad line region (BLR) radius and the AGN luminosity (the $R-L$ relation with scatter $\lesssim$0.2 dex; e.g. \citealt{kaspi2005relationship, bentz2009radius}), which enables the use of carefully chosen monochromatic luminosities as a surrogate for the BLR radius. Using this relation, a single observation could yield a proxy to the BLR radius, hence the single-epoch (SE) name. 
Despite their wide applicability and convenience, non-negligible systematic uncertainties affect these estimates, which are generally estimated to be of the order of 0.4--0.5 dex (see e.g. \citealt{shen2013mass} for a thorough discussion). An additional difficulty is caused by the complicated behaviour of the BLR, especially in the case of luminous sources approaching the quasar regime (e.g. \citealt{fries2024sdss}). Non-virial motions of the BLR and a possibly luminosity-dependent covering factor -just to mention some possible effects- ultimately led to even questioning the applicability of these calibrations for the quasar population at high redshift (\citealt{abuter2024dynamical, bertemes2025jwst}).

The new observational capabilities unlocked by the \textit{James Webb Space Telescope} (\jwst, \citealt{gardner2006james, gardner2023james}) are are dramatically improving our understanding of how the high-redshift Universe assembled. In particular, observations carried out with the NIRSpec instrument (\citealt{jakobsen2022near}) onboard \jwst revealed the properties of the interstellar medium (ISM) of the galaxies hosting the first generation of quasars at $z\gtrsim 3.5$ with unprecedented detail (e.g. \citealt{perna2023ga, loiacono2024quasar, marshall2023ga, liu2024fast, vayner2024powerful,parlanti2024ga, zamora2024ga, marshall2025jwst, bertola2025ga, perna2025ga}). Such observations revealed, for the first time in the rest-frame optical, extended emission in the ionised phase of the ISM, mostly in the previously inaccessible \oiii\ and \ion{H}{$\alpha$} lines.

Widespread outflows have been observed in different gas phases, such as ionised (in e.g. \citealt{bischetti2022suppression, bischetti2023fraction, yang2023spectroscopic, loiacono2024quasar}), and molecular (in e.g. \citealt{bischetti2019widespread, salak2024molecular, spilker2025direct}) across many sources at the epoch of reionisation. 
Several observations confirmed that the QSO-related outflow events expel significant amounts of ionised gas (larger than the measured star formation rates; SFR, see e.g. \citealt{venturi2023complex} and references therein), while even larger gas masses are expected to be ejected in the neutral phase (e.g. \citealt{cresci2023bubbles, davies2024jwst, deugenio2024fast, salak2024molecular}). Prolonged quasar outflow events, with similar energetics as those measured in these observations, are therefore expected to naturally deplete the host galaxies of their gas content (see e.g. \citealt{perna2018molecular, circosta2021super, bertola2024mol}, but see also \citealt{harrison2024observational}). The concomitant detection of such galaxy-scale outflows and the presence of the benchmark H$\beta$ line allowed for the first time to link the energetics of the quasar outflow to the well constrained black hole mass deep into the epoch of reionisation (e.g. \citealt{marshall2023ga, marshall2024ga, liu2024fast}).

In this paper, we present \jwst observations of the distant quasar ULAS J1342+0928 (\citealt{banados2018}). This source, at redshift $z=7.54$, represents the ideal laboratory to investigate the interplay between the central active nucleus and its environment at the epoch of reionisation. Crucially, the multi-wavelength campaigns that have targeted this object allowed for a complete characterisation of the broad-band properties of both the quasar and the host galaxy (see Sect. \ref{app:previous_studies}).

In Section 2 we present the new \jwst observations of ULAS J1342+0928 as well as previous observing campaigns. Sections 3 and 4 are dedicated to the analyses performed and the main results. We discuss our results in the context of high-redshift QSOs and AGN in Section 5 and collect our conclusions in Section 6. Throughout this paper, we adopt a flat $\Lambda$CDM cosmology with $H_0 = 70$ km s$^{-1}$ Mpc$^{-1}$, $\Omega_{\Lambda}$ = 0.7, and $\Omega_{m}$ = 0.3. Correlations are considered significant when yielding p-values lower than 0.01, corresponding to $\sim$2.6 $\sigma$ in the case of Normal approximation, under the null-hypothesis of non correlation.

\section{Data}
\label{sec:data}

\subsection{JWST/NIRSpec IFS observations of ULAS J1342}
\label{sec:data_red}
The quasar ULAS J1342+0928 (RA: 13:42:08.10, DEC: +09:28:38.36, $z=7.54$), ULAS J1342 for short, is the second most distant QSO known so far. It was first discovered in \citet{banados2018}, by exploiting the \ion{Ly}{$\alpha$} dropout technique, combining the Wide-field Infrared Survey Explorer (WISE, \citealt{wright2010wide}) ALLWISE catalogue (\citealt{cutri2013explanatory}), the United Kingdom Infrared Telescope Infrared Deep Sky Survey (UKIDSS, \citealt{lawrence2007}), and the DECam Legacy Survey (DECaLS; \citealt{blum2016decam}) photometry. Since then, several multi-band observational campaigns have targeted this object. We describe them in more detail in Appendix \ref{app:previous_studies}.

ULAS J1342 was observed as part of the NIRSpec GTO programme ‘Galaxy Assembly with NIRSpec IFS’ (GA-NIFS\footnote{\url{https://ga-nifs.github.io}}) under programme 1219 (PI: Nora L\"{u}tzgendorf). The target was observed on the 21st February 2023 with the Integral Field Spectrograph mode (IFS; \citealt{boker2022near}) using a medium cycling pattern of 6 dithers and a total integration time of 3~h with the high-resolution grating/filter pair G395H/F290LP covering the wavelength range 2.9--5.3 $\mu$m with a spectral resolution R $\sim$ 1900--3600 (\citealt{jakobsen2022near, boker2022near}). A 1D spectrum extracted from this cube centred at the quasar location was also employed in \citet{christensen2023metal}, in addition to the G140H/F070LP and the G235H/F170LP fixed slit data, to cover the \Hb--\oiii region. Additionally, PRISM/CLEAR observations (R~$\sim$~30--330) in the 0.6--5.3 $\mu$m wavelength range, with a medium cycling pattern of 6 dithers and a total integration time of 1~h, were also performed. Although delivering a lower spectral resolution, PRISM data yield a superior sensitivity, key for detecting faint extended features.

We retrieved the raw data from the the Barbara A. Mikulski Archive for Space Telescopes (MAST) and adopted a modified version of the \jwst pipeline (version 1.8.2), under the Calibration Reference Data System context jwst\_1068.pmap. Our customisation of the standard calibration pipeline resulted in improved data quality. Most of these improvements are described in \citet{perna2023ga}. Here we briefly highlight the major additions. We corrected the 1/f noise via a polynomial fit and removed regions affected by failed opening of the MSA shutters by masking the corresponding pixels. We cleaned individual exposures from the presence of cosmic rays by taking advantage of an algorithm akin to \textsc{LACOSMIC} (\citealt{van2001cosmic}) to prune outliers in the individual 2D exposures. In short, as explained in \citet{deugenio2024fast}, we measured the distribution of the derivative values along the dispersion direction, normalised it by the maximum of the flux (or three times the RMS noise, if this was larger) and rejected the pixels above 95$\rm^{th}$ percentile of the distribution. We finally combined the cube using the `drizzle' method, producing a cube with $0\farcs05$ spaxels. Because of the brightness of the point source QSO, as well as the under-sampling of the PSF (see e.g. \citealt{smith2007spectral, law20233d}), the final cube was affected by sinusoidal-type patterns (often referred to as `wiggles') at the level of single spaxels. We corrected for this instrumental effect adopting the same recipe described in \citet{perna2023ga}. Sky background subtraction was performed by subtracting from each spaxel the smoothed spectrum integrated in a source-free region of the field of view.

\begin{figure}[h!]
\centering
\includegraphics[width=\linewidth,clip]{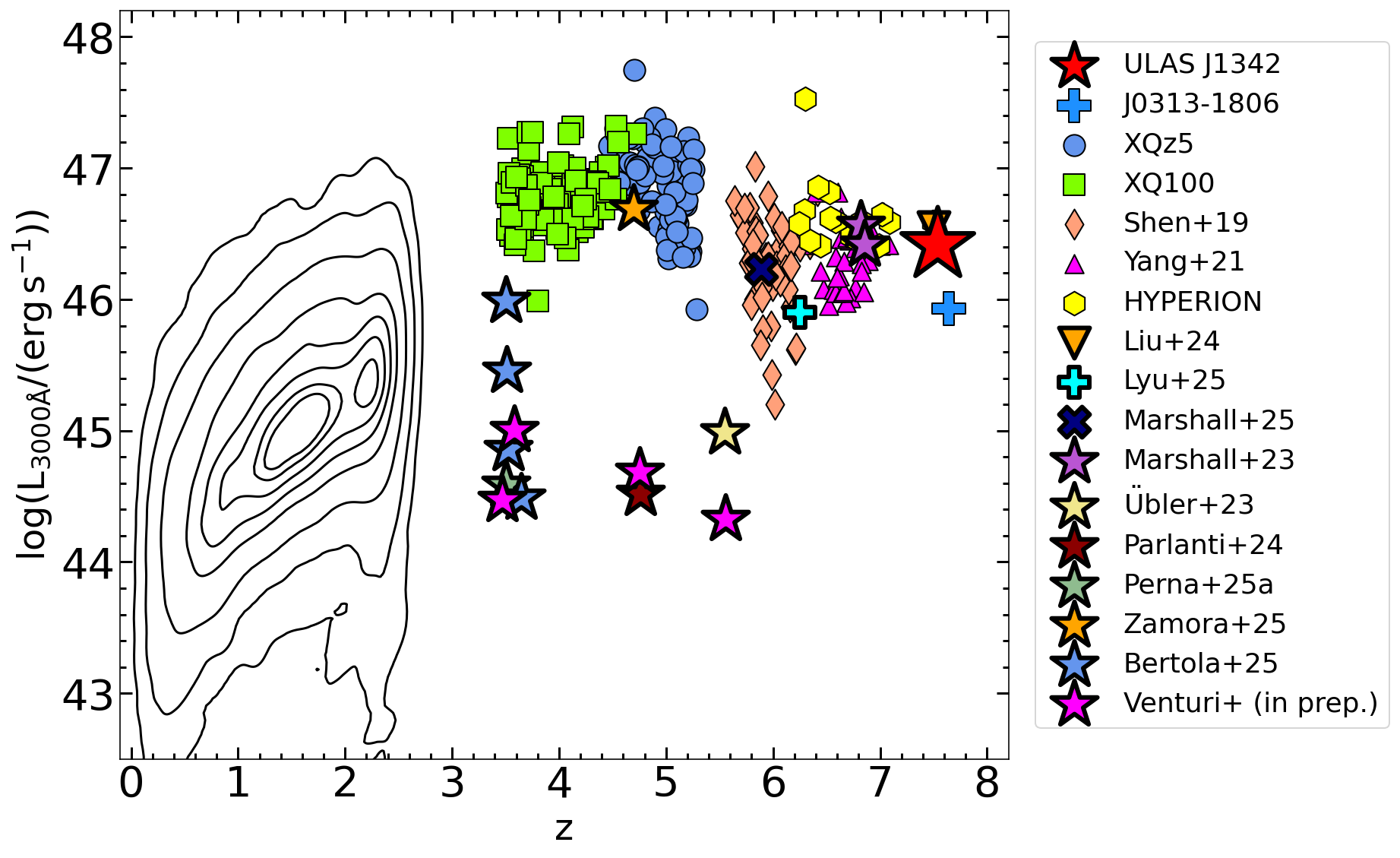}
\caption{ULAS J1342 in the context of other QSOs surveys. The most distant quasar known so far J0313-1806 (\citealt{wang2021luminous}) has been shifted downwards by 0.5 dex for graphical purposes. Some objects are shared among multiple surveys. Star-shaped symbols mark objects observed within the GA-NIFS programme \citep{ubler2023ga, marshall2023ga, marshall2024ga, parlanti2024ga, zamora2024ga, perna2025ga, bertola2025ga}. Thick-edged symbols mark other objects observed with JWST NIRSpec IFU (\citealt{liu2024fast, lyu2025fading, marshall2025jwst}).}
\label{fig:z_L3000}
\end{figure}

We show the redshift and 3,000\AA\, luminosity of ULAS J1342 in the context of other QSOs surveys at different redshifts in Fig. \ref{fig:z_L3000}. There, we also highlight the parameter space explored by other type 1 AGN within the GA-NIFS programme (see Sect. \ref{sec:data_red}). We show the SDSS DR16 sample (\citealt{wu2022catalog}) as black contours, and the XQz5 QSOs (\citealt{lai2024xqz5}), the XQ100 sample (\citealt{lopez2016}), the QSOs from \cite{shen2019gemini}, the sample from \cite{yang2021probing}, and those from the HYPERION sample (\citealt{zappacosta2023hyperluminous}) with other symbols according to the legend.

\section{Methods}
\label{sec:methods}

\subsection{Spectral fit of the high-resolution nuclear spectrum}
\label{sec:blr_fit}

\begin{figure*}[h!]
\centering
\includegraphics[width=\linewidth,clip]{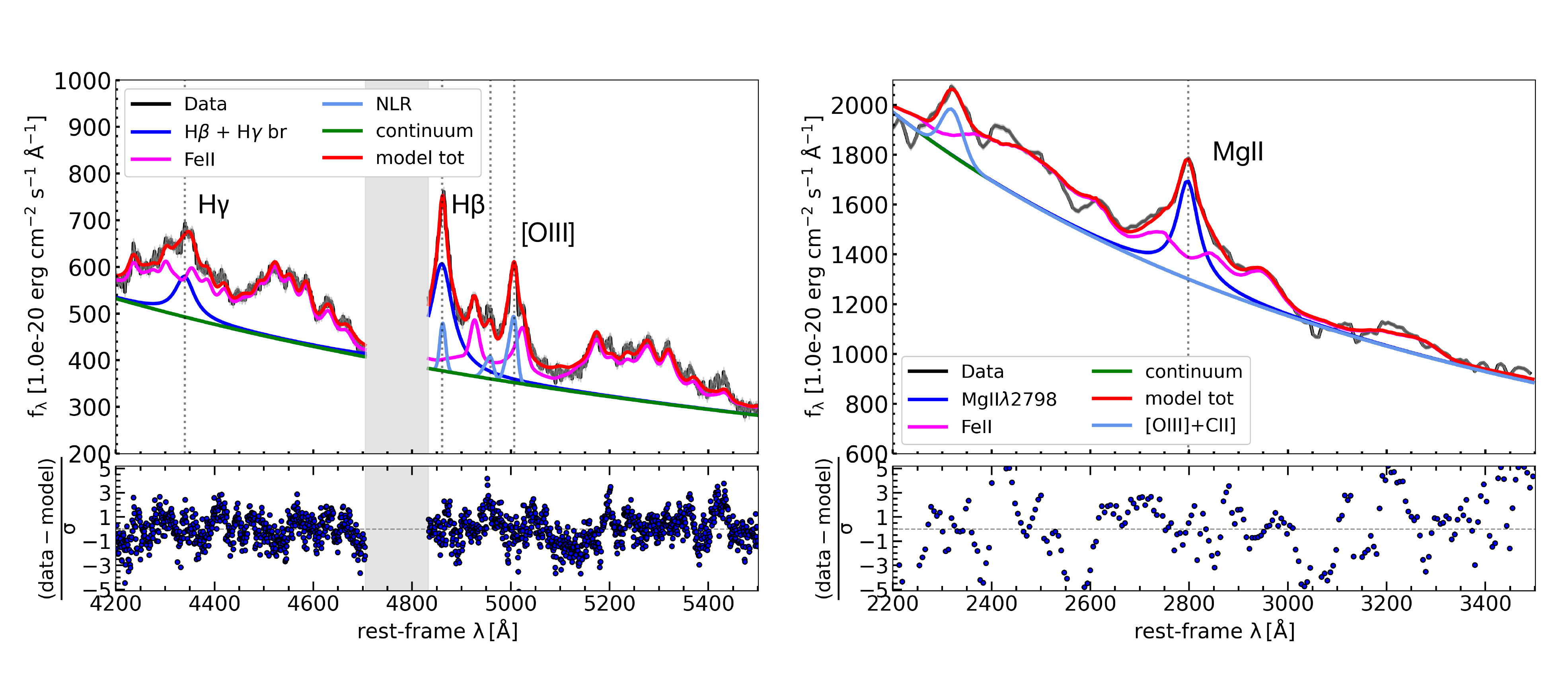}
\caption{\textit{Left:} Spectral fit of the integrated R2700 spectrum extracted from a 6$\times$6 spaxels ($0\farcs3$ radius) region centered at the QSO location. Vertical dotted lines mark the position of the main emission lines. \textit{Right:} Same as in the left panel for the prism spectrum around the \ion{Mg}{ii} emission line. In both images, the fit components are colour-coded according to the legend. Residuals are shown in the bottom panels.}
\label{fig:spec_fits_BLR}
\end{figure*}

In the left panel of Fig.~\ref{fig:spec_fits_BLR} we show the G395H/F290LP integrated spectrum extracted from a circular region with a $0\farcs3$ radius, centred on the QSO location, covering the rest-frame \Hb--\oiii region (shown in magenta in Fig. \ref{fig:o3_map_spectra}). In the extraction, with the aim of producing a more realistic estimate of the flux uncertainties, we rescaled the formal uncertainty on the integrated spectrum based on the "ERR" extension using the flux standard deviation in small ($\sim$20--30 \AA) continuum windows free from strong emission lines. This allowed us to take into account the correlations induced by the size of the PSF relative to the spaxel size (see e.g. \citealt{ubler2023ga}).

We performed the spectral fit of the integrated spectrum employing a custom-made Python code, based on the IDL \textsc{MPFIT} package \citep{Markwardt2009}, which takes advantage of the Levenberg-Marquardt technique \citep{more1978levenberg} to solve the least-squares problem. The model is composed of a power law continuum (associated to the accretion disc), broad (from the BLR) and narrow (from the narrow-line region, NLR) emission lines as well as the optical \ion{Fe}{ii} pseudo continuum (also from the BLR). 

The broad ($\rm FWHM>1,000 \, km \, s^{-1}$) emission lines (\Hb and \ion{H}{$\gamma$})\footnote{Here the term broad lines refers to those arising from the BLR, rather than to the broad components indicating the presence of an outflow.} were modelled by adopting different Lorentzian profiles which proved more effective than Gaussian line profiles in reproducing the line shapes. We tied the kinematics of the NLR components (i.e. \oiii$\lambda\lambda$4959,5007 and narrow \Hb) and the \oiii$\lambda$5007/\oiii$\lambda$4949 flux ratio was fixed to its theoretical value of 3 (\citealt{osterbrock2006astrophysics}). We also included outflow components in both \oiii and \Hb, and constrained their kinematics to be the same. Both the systemic and the outflow components were modelled as Gaussian lines.
ULAS J1342 presents a fairly complex, challenging to model, \ion{Fe}{ii} emission. We reproduced it by employing a linear combination of twenty spectral templates derived within the \textsc{Cloudy} environment (\citealt{ferland2013}). These models are then convolved with a Gaussian profile (the same for all the models as they are expected to represent co-spatial emission), spectrally shifted, and weighted during the fitting process. The parameters ultimately fitted for the \ion{Fe}{ii} templates are then the velocity dispersion and the velocity shift of the Gaussian kernel, as well as the weights of each template. We paid particular attention to this modelling, even adding a couple of additional broad components to faithfully reproduce single \ion{Fe}{ii} multiplets, as the BLR template resulting from this analysis was then used for the spaxel-by-spaxel PSF subtraction procedure which we describe in Sect. \ref{sec:spat_specfit}.

As the nuclear spectrum is integrated from a finite aperture, the fluxes evaluated are smaller than the intrinsic ones because of aperture losses. We therefore accounted for this effect by estimating the aperture correction using the strong broad optical \Feii which is expected to be produced only on unresolved scales\footnote{Given that the wavelength ranges analysed here are rather small for both the \Hb--\oiii and the \ion{Mg}{ii} complexes, we applied a fixed (i.e. not wavelength-dependent) aperture correction for each spectral region.}. Here, we refrained from using the \Hb because of the detector gap falling on the blue side of the line. We compared the flux of the integrated \Feii in the nuclear aperture ($0\farcs3$) and in a much wider aperture virtually including the whole PSF ($1\farcs5$; see e.g. \citealt{marshall2023ga, liu2024fast}). We found that the correction needed to account for the aperture losses is about $16\%$, the same amount quoted in works targeting bright QSOs at similar $z$ (\citealt{liu2024fast}). We applied the same recipe to derive the aperture correction for the low resolution data, which amounts to $\sim 12\%$. The redshift derived from the rest-frame set of NLR lines is $z=7.535$, and we will use this value as a reference in the following.

We estimated the uncertainties on the best-fit parameters by adopting a Monte Carlo procedure. In brief, we created 100 mock spectra by adding noise in each spectral channel to the starting spectrum. The noise was drawn from a Gaussian distribution, with wavelength-dependent amplitude equal to the nominal uncertainty in each spectral channel. We then proceeded by fitting the mock spectra adopting the same model employed for the observed spectrum and computed the distribution of the best fit parameters. The final uncertainties on the best-fit parameters were estimated as the standard deviations of these distributions after applying a 3$\sigma$ clipping to prune outlier values.

We report the relevant parameters derived from the fit together with the relative uncertainties in Table \ref{tbl:fit_pars}.

\begin{table}[h!]
\centering
\setlength{\tabcolsep}{5pt}
\caption{Parameters derived from the spectral fitting to the nuclear integrated spectrum.}
\begin{tabular}{c c}
 \hline \noalign{\smallskip}
 Parameters & Value \\
 (1) & (2) \\

 \hline \noalign{\smallskip}
 $\rm \log(L_{H\beta})$      & $44.18 \pm 0.01$ \\ 
 $\rm FWHM_{H\beta}$         & $3300 \pm 440$   \\ 
 $\rm \log(L_{5,100\AA})$     & $46.13 \pm 0.01$ \\ 
 $\rm EW_{FeII}/EW_{H\beta}$ & $1.11 \pm 0.03$    \\ 
 $\rm \alpha_{\lambda,opt}$  & $-2.35 \pm 0.04$   \\ 

 \hline

 $\rm \log(L_{MgII})$      & $44.39 \pm 0.01$ \\ 
 $\rm FWHM_{MgII}$         & $3360 \pm 300$   \\ 
 $\rm \log(L_{3,000\AA})$   & $46.42 \pm 0.01$ \\ 
 $\rm EW_{FeII}/EW_{MgII}$ & $3.41 \pm 0.10$  \\ 
 $\rm \alpha_{\lambda,UV}$ & $-1.73 \pm 0.01$ \\

 \hline
\end{tabular}

\tablefoot{The \Hb luminosity (in $\rm erg \, s^{-1}$) and FWHM only account for the broad component. FWHMs of both high- and low-resolution data have been corrected for instrumental line spread function, using the available g395h and the PRISM dispersion curves, and are in units of $\rm km\, s^{-1}$. All of the parameters for the H$\beta$ have been measured on the high-resolution spectrum, while those for the \ion{Mg}{ii} line are measured from the prism as explained in the text. The equivalent widths (EWs) of optical and UV \ion{Fe}{ii} have been computed by integrating the best fit models between 4434--4684~\AA\, and 2200--3090~\AA, respectively. Here we enforced an uncertainty of 0.01 dex for monochromatic luminosities and 0.01 for continuum slopes ($\rm \alpha_{\lambda,UV}$, $\rm \alpha_{\lambda,opt}$).}
\label{tbl:fit_pars}
\end{table}

\subsection{Spectral fit of the low-resolution nuclear spectrum}
The spectral properties tracing the kinematics of the nuclear region are best described by the high resolution data. However, the prism data, due to the higher sensitivity to high-EW, spectrally broad lines are well suited to detect the diffuse emission, mostly seen in \oiii (see also Sect. \ref{sec:diffuse_o3}). For this purpose, we need to correct for the nuclear emission polluting nearby spaxels due to the finite size of the \jwst PSF. To this end, we independently performed a spectral fit of the \Hb--\oiii nuclear spectrum in the prism data (see Fig. \ref{fig:hb_fit_prism}), adopting the same method employed for the high resolution spectrum. The best fit model of the broad lines was then adopted as a BLR template and used to subtract the PSF contamination in the spaxel-by-spaxel fit (Sect. \ref{sec:spat_specfit}).

In addition, further information about the BLR properties of the source can be gathered by analysing the spectral region covering the rest-frame \ion{Mg}{ii}$\lambda$2798 line, which was already covered with ground-based IR spectroscopy (\citealt{onoue2020no}). We employed the same routine described above to fit the spectral region around the \ion{Mg}{ii} line. In this case, we included in the model only the broad \ion{Mg}{ii}$\lambda$2798 line, the narrow \oiii$\lambda$2321 and \ion{C}{ii}]$\lambda$2324 blended together. We also added a power-law continuum between 2,100\AA--3,500\AA\ and the UV \ion{Fe}{ii} which, given the resolution, blends into a pseudo-continuum. Again, the \ion{Fe}{ii} was modelled using the \textsc{Cloudy} templates already mentioned in Sect. \ref{sec:blr_fit}. We corrected the line width for the prism line spread function, which at the \ion{Mg}{ii} wavelengths ($\sim 2.4~ \mu$m in the observed frame) has a resolution of R$\lesssim$100. The intrinsic line FWHM was then computed by subtracting in quadrature the instrumental broadening from the observed FWHM. We show the result of the fit of the \ion{Mg}{ii} region in the right panel of Fig.~\ref{fig:spec_fits_BLR}.

\subsection{Accretion disc modelling}
\label{sec:ad_modelling}
Accretion disc modelling offers a viable alternative to estimate the parameters of the accretion flow powering the quasar emission in ULAS J1342 (e.g. \citealt{malkan1983ultraviolet}). This method relies on a relatively wide spectral coverage, including the region close to the peak of the emission, generally falling in the UV for SMBHs, and the continuum to be measured in emission line-free windows. In exchange, it offers favourable aspects, such as a sounder understanding of the physics underlying the optically thick emission of the accretion disc, as compared to the more loosely constrained kinematics of the BLR. Additionally, accretion disc models naturally provide a self-consistent estimation of $\rm M_{BH}$ and $\rm L_{bol}$. In particular, this approach has already been undertaken for this source in \citet{campitiello2019black} using ground-based spectroscopy mostly covering the spectral region between $\sim 1,400$--$3,000$ \AA. There the authors, adopting different accretion disc models, estimated $\rm M_{BH}$ to be in the range $10^{8.9-9.2}$ \Msun. 
The nuclear prism spectrum is ideal to perform such analysis. The PRISM/CLEAR data have indeed a high-enough resolution to select the continuum emission in emission line-free regions, while, at the same time, offering a wide spectral coverage from the \ion{Ly}{$\alpha$} up to \Hb wavelengths. Additionally, in objects as luminous as ULAS J1342 the turnover of the optical continuum at $\lambda \gtrsim 4,000$ \AA, generally associated with the host galaxy (e.g. \citealt{vandenberk2001}), is not observed, therefore allowing for a wider continuum baseline.

\begin{figure*}[h!]
\centering
\includegraphics[width=0.7\linewidth,trim=0mm 0mm 0mm 0mm,clip]{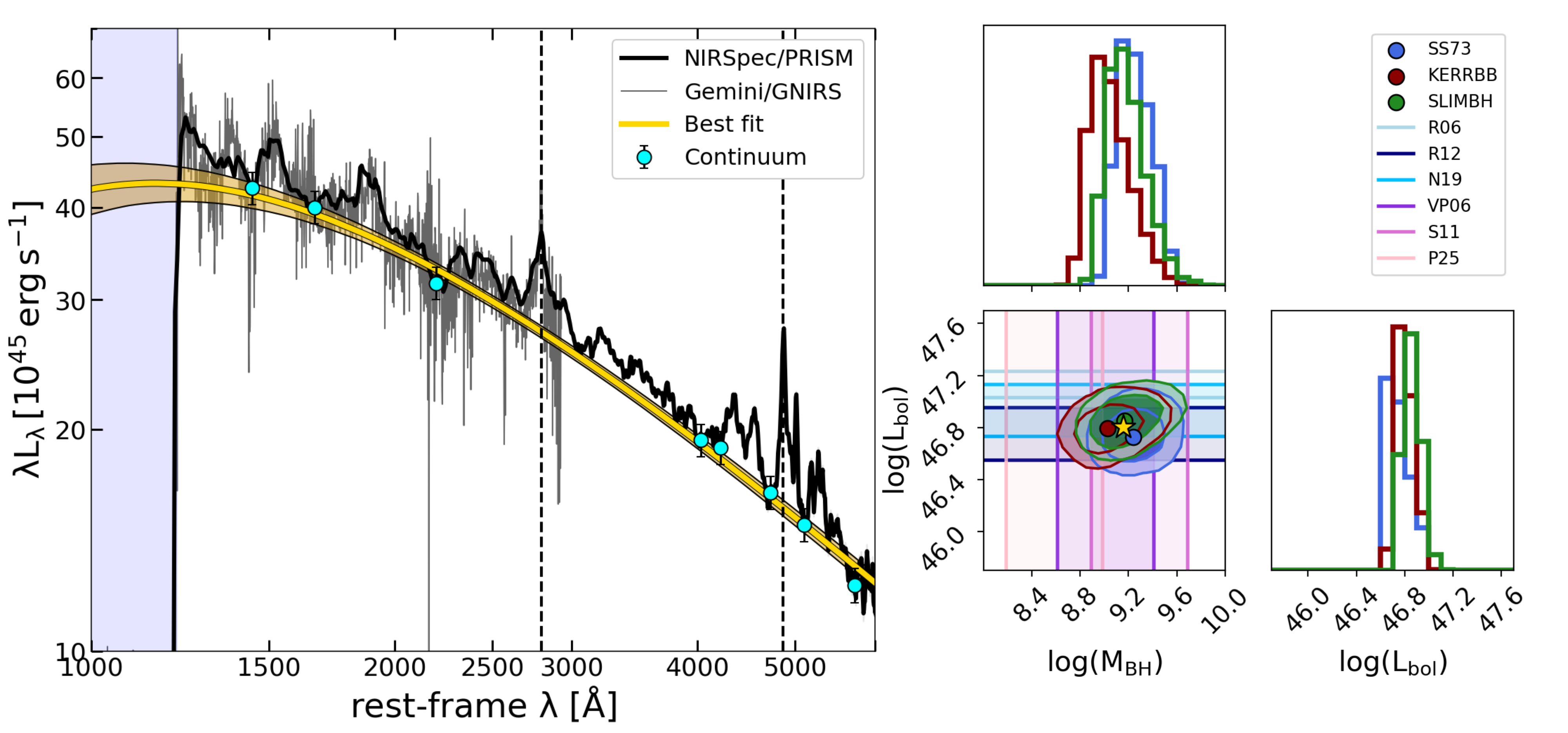}
\caption{Accretion disc modelling of the nuclear prism spectrum (black). We also show the Gemini/GNIRS spectrum from \citet{onoue2020no} overlapping in the spectral region below 3,000 \AA. The continuum points employed in the fit are marked as cyan dots. The gold line marks the best fit, while the shaded band represents the 16$\rm^{th}$-84$\rm^{th}$ percentiles of the distribution of 1000 randomly extracted best fit models from the joint posterior distribution. The vertical dashed lines mark the broad emission lines (\ion{Mg}{ii}$\lambda$2798, \Hb) used to compute the black hole mass via the single-epoch virial relations. The blue shaded area highlights the region bluewards of the \ion{Ly}{$\alpha$} affected by inter-galactic medium (IGM) absorption, and therefore excluded in the fit. The inset panel shows the joint posterior distributions of $\rm M_{BH}$ and $\rm L_{bol}$ together with the independent estimates coming from optical bolometric corrections (R06, R12, N19, see Sec. \ref{sec:accretion_parameters}) and the masses from virial calibrations (VP06, S11, P25, see Sec. \ref{sec:accretion_parameters}). The coloured points mark the median values of the posterior distributions which are combined to estimate the best-fit value (yellow star)}.

\label{fig:ad_fit}
\end{figure*}

For the accretion disc modelling, we isolated the continuum emission in small spectral windows (spanning $\sim 20$--$30$ \AA, see Fig.~\ref{fig:ad_fit}) avoiding strong emission lines as well as the small blue bump between $\sim$2,200--4,000 \AA\, (\citealt{grandi1982}), which is made of the blend of the \ion{Fe}{ii} pseudo-continuum and the Balmer continuum. The uncertainty on each continuum point was evaluated by simple error propagation, and is of the percent order. However, this approach would not take into account the flux calibration uncertainty, which could dominate the uncertainty budget. The required accuracies for the flux calibration of \jwst spectroscopy are of the order of 10\%–15\% (\citealt{gordon2018end}). On-flight measurements on a relatively small sample of standard stars proved an average accuracy of $\lesssim 2\%$ (\citealt{boker2023orbit}) employing different filter/grating combinations. In order to retain a conservative approach, we introduced a factor of 5\% of the flux in the uncertainty budget. With the purpose of showcasing the capabilities of JWST/NIRSpec with respect to ground-based NIR facilities, in Fig.\ref{fig:ad_fit} we also overlay the Gemini/GNIRS long-slit spectrum described in \citet{onoue2020no}. The latter dataset was obtained with a total exposure of 9 h and reaches a median S/N at wavelengths larger than the \ion{Ly}{$\alpha$} of $\sim$30. As a comparison, our low-resolution dataset, in the same wavelength range, reaches a S/N$\sim$130 with just 1 h of exposure.

In order to mitigate our lack of knowledge about the accretion state of the quasar disc, we adopted three complementary models with slightly different underlying physical assumptions. In more detail, we used a standard non-relativistic \citet{shakura1973black} geometrically thin and optically thick accretion disc. In order to allow for different mass-to-light conversion efficiencies ($\epsilon$), we included the innermost stable orbit of the disc ($\rm R_{ISCO}$) as a free parameter. In the Newtonian limit $\rm \epsilon=R_{ISCO}/2R_G$, with 2$\rm R_{G}$ being the Schwarzschild radius. Moreover, the inclination angle of the line of sight (LoS) with respect to the accretion disc axis was also included as a free parameter, although the Type 1 nature of this quasar limits the possible viewing angles in the range $\rm \theta \sim [0^{\circ}$--$70^{\circ}]$. For this model, the parameters ultimately left free to vary during the fit were the black hole mass, the disc luminosity, the innermost disc radius, and the inclination angle.

As complementary approaches, we also fitted the data employing the {\tt KERRBB} model, a multi-temperature blackbody model for a thin, steady-state accretion disc around a Kerr black hole (\citealt{li2005multitemperature}), and the {\tt SLIMBH} model, which also accounts for a thickening of the accretion disc at high accretion rates (\citealt{skadowski2009slim}). Both of these models are included in \textsc{ciao} (\citealt{fruscione2006ciao}), the modelling application within the \textsc{sherpa} environment (\citealt{siemiginowska2024sherpa}), which takes advantage of the \xspec (\citealt{arnaud1996xspec}) library of models. In particular, we used \textsc{sherpa}/\textsc{ciao} version 4.17.0. For these two latter models, we fitted for the black hole dimensionless spin ($a$) instead of the innermost disc radius, included among the \xspec model parameters. The fit was then performed adopting a full Monte Carlo Markov Chain (MCMC) Bayesian approach implemented through the \textsc{emcee} python package (\citealt{emcee}). Since we do not know a priori which of the proposed models can yield the most realistic description of the accretion disc state, we opted for a Bayesian Model Averaging (e.g. \citealt{yao2018BMA}) under the simplifying hypothesis of equi-probability of the models. We show the result of the accretion disc modelling in Fig. \ref{fig:ad_fit}. As a further check for the reliability of our fitting routine, we repeated the fit employing the ``BADFit'' routine (\citealt{lai2023characterising,samuel_lai_2023_7772748}) which implements in a Bayesian framework both the {\tt KERRBB} and the {\tt SLIMBH} models. Both the black hole mass and the disc luminosity, estimated with this alternative recipe, proved consistent with our own implementation within the uncertainties, which are of the same order of magnitude as ours.

We note that here we did not take into account the possibility of significant dust extinction along the line of sight. We cannot probe it directly using the narrow Balmer lines decrement, because of the lack of the \ion{H}{$\alpha$} in our data, and the faintness of the \ion{H}{$\gamma$}. Furthermore, such ratios would only constrain the extinction in the NLR scales. On a general ground, the steep continuum observed argues against the possibility of significant dust extinction. For instance, the quasar spectral template of \citet{selsing2016x}, purposefully made of a sample including remarkably luminous blue quasars (i.e. unobscured) between $1<z<2$, exhibits a spectral slope $\alpha_{\lambda}=-1.70$ in the range $\sim 1,450$--$5,500$ \AA. As a comparison, ULAS J1342 has a spectral slope $\alpha_{\lambda}=-1.77$ in the same spectral range, whereby the presence of dust reddening would make this slope shallower. A similar argument can be drawn by considerations about the optical slope. The slope of the continuum measured between $\sim 4,200$--$5,500$ \AA\, is $-2.35 \pm 0.04$ (see Table \ref{tbl:fit_pars}), consistent with the expectations for an optically thick accretion disc in the $\nu L_{\nu} \sim \nu^{4/3}$ region, corresponding to $\alpha_{\lambda}\sim -2.33$. This does not leave much room for either dust extinction and/or host galaxy contamination at optical wavelengths. Lastly, we also mention that complementary information coming from X-ray data argue in favour of a low extinction delivered by dust-free gas, if any. The photon index measured in the two observing campaigns ranges between $\Gamma=1.9$ (\citealt{banados2018chandra}) and $\Gamma=2.9$ (\citealt{zappacosta2023hyperluminous,tortosa2024hyperion}). Conversely, smaller photon indices ($\Gamma \lesssim 1.7$, see e.g. \citealt{merloni2014incidence}) are found in obscured sources where gas extinction mostly absorbs the soft part of the X-ray spectrum ($\lesssim 2$ keV), effectively producing a slope flatter than the intrinsic. Lastly, we note that, if present, reddening would cause our SE and AD \mbh estimates to diverge: SE \mbh rely on the measurement of some continuum or line measurement, and would therefore be underestimated in case of reddening. On the other hand, in the case of a non-greybody like extinction curve, dust extinction makes the observed SED redder (i.e. colder), hence pushing the AD fit \mbh to larger values. Not observing such discrepancy, with \mbh being consistent (within the systematic uncertainties) between the two methods, also argues against the presence of significant reddening. Notwithstanding all these considerations, we also explored the effect of reddening by directly including it into our AD modelling in Appendix \ref{app:extinction}.

\subsection{Spatially resolved spectral fits}
\label{sec:spat_specfit}
With the aim of characterising the spatially resolved properties of the gas in the close environment of ULAS J1342, we performed a spaxel-by-spaxel fit of the spectroscopic data. For this purpose we employed the same routines described in Sect. \ref{sec:blr_fit} for each spatial element of the NIRSpec IFU for both  the high- and the low-resolution cubes. In order to account for the extent of the PSF, we created a template made of the BLR components and the continuum from the best-fit model shown in Fig.~\ref{fig:spec_fits_BLR}. We added this component, whose normalisation was left free to vary (e.g. \citealt{carniani2015ionised, venturi2018magnum, marasco2020galaxy}), to the narrow Gaussian lines and a local second degree polynomial continuum to model the \Hb--\oiii complex on a spaxel-to-spaxel basis. In particular, we used up to two components for both \Hb and \oiii in the high resolution data (corresponding to a narrow line and a broader outflow component), while only one was included to reproduce the prism cube.

\section{Results}

\subsection{Tracing the diffuse [\ion{O}{iii}] emission}
\label{sec:diffuse_o3}

\begin{figure*}[h!]
\centering
\includegraphics[width=0.9\linewidth, trim=0mm 14mm 0mm 14mm,clip]{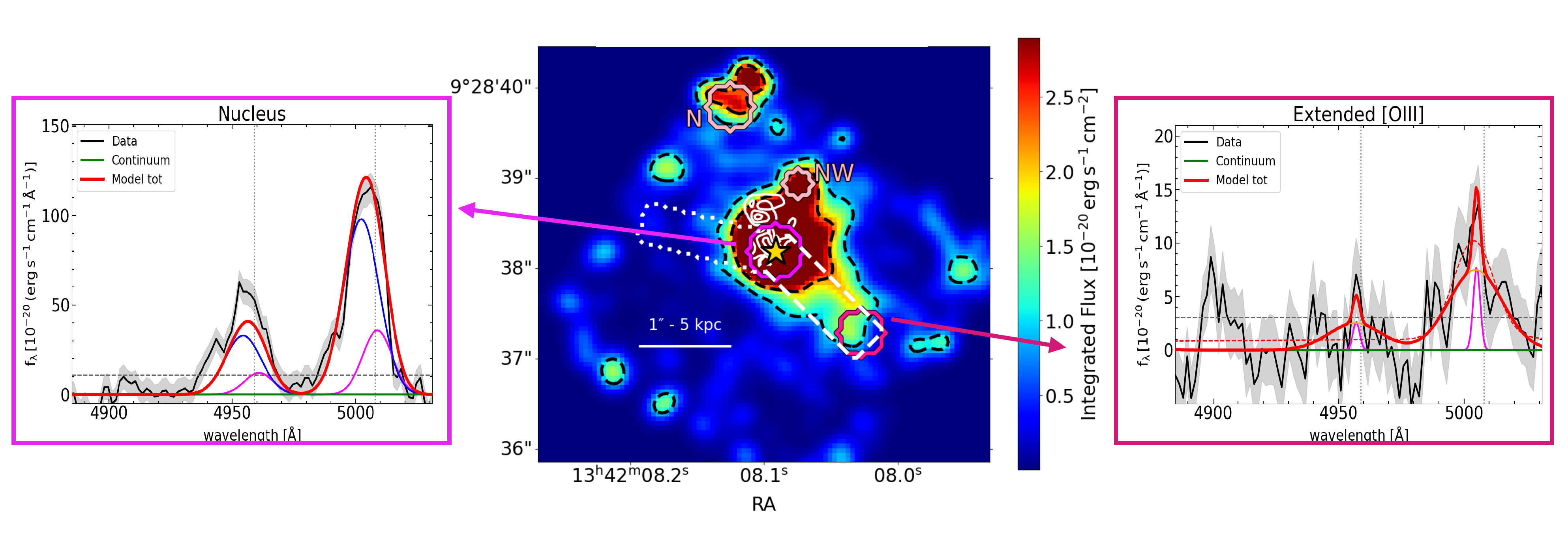}
\caption{\oiii flux map from the prism BLR- and continuum-subtracted cube. North is up, East is left. The star marks the QSO location, while the white solid contours show the overlaid ALMA [\ion{C}{ii}] emission. The dashed black contours mark the \oiii emission at 3 and 7 $\sigma$, respectively. The circles show the extraction regions of the \oiii spectra from the R2700 cube, which are labelled accordingly. The white dashed box highlights the direction along which the extended \oiii surface brightness profile shown in Fig. \ref{fig:O3_sbp} was calculated. The dotted one has the same meaning for the \oiii control direction. In the inset panels, we show the R2770 spectra together with the best fit models for the nucleus and the \oiii halo, while those for the N and NW \oiii emitters are presented in Appendix \ref{app:o3emitters_spectra}. In the spectra, the vertical dotted lines mark the expected \oiii wavelengths at the systemic redshift. The horizontal dashed line shows the RMS. The magenta Gaussian lines in both panels show the narrow rest-frame component, while other Gaussian components reproduce the outflows. In the nuclear spectrum, the flux excess in \oiii$\lambda$4959, not well reproduced by the fit, is mostly due to \Feii contamination not perfectly accounted for by the BLR best fit model (see Fig.~\ref{fig:spec_fits_BLR}). In the spectrum extracted in the extended \oiii halo (top-right panel), we show both the best-fit models, marking with the solid red line the model with one broad and one narrow components and with the dotted one that with only one broad component. The image has been smoothed using a Gaussian kernel 1 pixel wide.}

\label{fig:o3_map_spectra}
\end{figure*}

In Fig.~\ref{fig:o3_map_spectra} we present the BLR-subtracted low-resolution datacube integrated across the \oiii wavelengths. Several noticeable features clearly stand out. In addition to the nuclear \oiii component, we report the presence of multiple structures. The most striking feature is an elongated \oiii halo extending for $\sim 1\farcs4$ towards the SW. We discuss in greater detail the properties of this region in Sect. \ref{sec:extO3}.

In addition, we detected two \oiii emitters with a clumpy morphology within the field of view, respectively north and north-west of the QSO. We present the high-resolution spectra of all of the circular regions highlighted in Fig.~\ref{fig:o3_map_spectra} in the surrounding panels or in Appendix \ref{app:o3emitters_spectra}. One of these clumpy \oiii-emitting regions is located at $\sim 0\farcs7$ north-west of the central QSO (labelled as NW), to which it appears to be linked by an ionised gas bridge, possibly hinting to an on-going interaction between these two systems.

The high-resolution spectrum of the NW clump, shown in Fig. \ref{fig:NW_clump}, revealed the presence of two separate sets of emission lines which appear to be consistent with being two distinct kinematic components of the \oiii$\lambda\lambda 4959,5007$ doublet. Assuming the redshift of the QSO to be traced by the NLR at $z=7.535$, one of the two \oiii doublets is redshifted by 140$\pm$15 km s$^{-1}$, while the other by 700$\pm$10 km s$^{-1}$. The FWHM of the most redshifted \oiii emitter is $\sim$200 km s$^{-1}$, while that of the other one is $\sim$300 km s$^{-1}$. Being the spectral resolution FWHM $\sim$ 110 km s$^{-1}$, we consider only the set of lines with FWHM equal to $\sim$ 300 km s$^{-1}$ as spectrally resolved, while the other as marginally spectrally resolved. These two kinematical components are not resolved in the low-resolution cube. In this framework, the two sets of lines would correspond to two physically separated systems, with the redshifted one being at $z$=7.557. However, due to its compact morphology in the high-resolution cube, and the \oiii being marginally resolved, we also mention the possibility that the most redshifted set of lines detected in this region could actually be an artifact introduced during the data reduction stage. None of our major conclusions would be affected by such an occurrence.

Moreover, we detected another \oiii emitter at $\sim 1\farcs8$ north of the QSO location (dubbed as N in Fig.~\ref{fig:o3_map_spectra}, see Fig. \ref{fig:N_clump}). For this region, we avoided performing the extraction close to the IFU edge as it would result in a noisier spectrum. Also in this case, the redshift of this structure is consistent with that of the QSO. The line width is close to the resolution limit ($\sim$200 km s$^{-1}$). While in the NW region the detection of the rest-frame UV continuum in the low-resolution data was hampered by PSF extent which smears the QSO continuum on larger scales, for the N clump we measured a rest-frame UV luminosity between 2,500--3,000~\AA\ of $\rm (5\pm0.2)\times10^{42} \, erg \, s^{-1}$. Both clumps are also detected in [\ion{O}{ii}]$\lambda3728$, while this is not the case for the extended \oiii halo, as we discuss in Appendix \ref{app:o2map}.

\subsubsection{The extended \oiii halo}
\label{sec:extO3}
\begin{figure}[h!]
\centering
\includegraphics[width=\linewidth,clip]{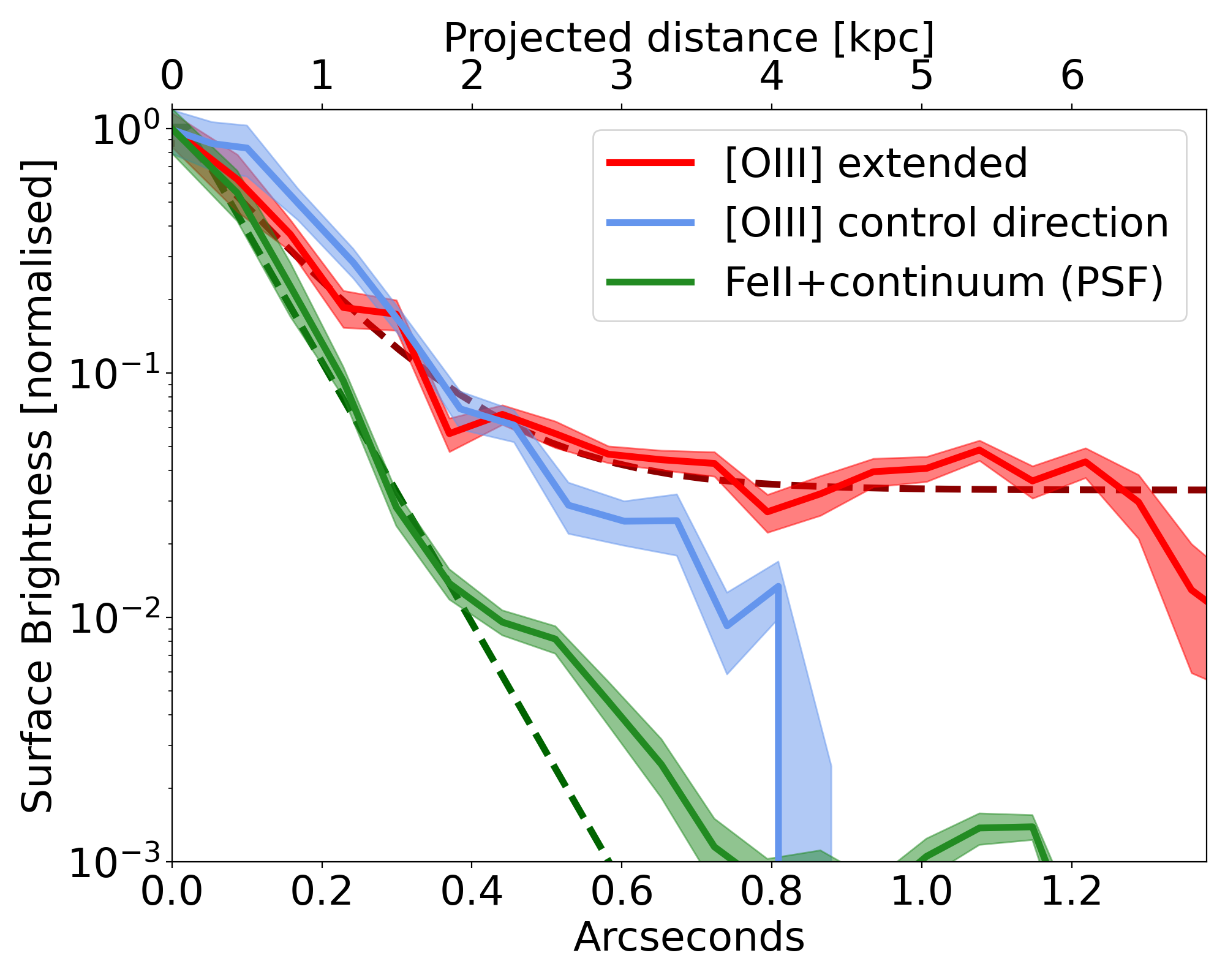}
\caption{Surface brightness profile of the extended \oiii emission (red) and PSF (green), as traced by the emission between 4464--4684 \AA. The \oiii emission evaluated across the same wavelengths along a control direction is also shown in cyan. All the surface brightness profiles are normalised to their peak value. An exponential profile plus a constant term fit to the profiles are shown as dashed lines.}
\label{fig:O3_sbp}
\end{figure}

In order to quantify the extent of the \oiii elongated component, and compare it to the PSF size, we produced the \oiii surface brightness profile along the SW direction employing the aperture shown as a white dashed rectangle in Fig.~\ref{fig:o3_map_spectra}. For this purpose, we integrated the low-resolution BLR-subtracted cube within $\pm 1,500$ km/s from the rest frame \oiii peak wavelength, after having subtracted a local continuum (only representing a marginal improvement) to account for a non optimal continuum subtraction in the spaxel-by-spaxel fitting. The PSF surface brightness profile was computed from the emission of the \ion{Fe}{ii} bump between 4464--4684 \AA\, which arises from the unresolved parsec scales of the QSO's BLR. The radial profiles of \oiii and PSF are shown in Fig. \ref{fig:O3_sbp}. We also extracted the \oiii profile from another random direction avoiding the companions (NE dotted rectangle in Fig.~\ref{fig:o3_map_spectra}) to study its morphology in the nuclear region and compare it with the kpc-scale SW plume. Analysing the profile of the \oiii in the two directions, we note that the surface brightness profile includes a compact and symmetric component extending up to $\sim 0\farcs4$ (i.e., 2 kpc) and the extended component up to $1\farcs 4$ (i.e., 7 kpc) toward the SW.
Empirically, a simple exponential brightness profile plus a constant well describes the observed profile. By comparing the width of the \oiii profile with that of the PSF it is clear that i) the \oiii is spatially more extended than the PSF and ii) this is true also in the central $0\farcs 4$, both along the extended halo, as well as along the control direction (whose profile mostly overlaps with the extended one below $\rm 0\farcs 4$, though their extraction region not being co-spatial). This suggests the presence of a spherically symmetric bright nuclear component, and a low-surface brightness one extending toward the SW.
With the aim of estimating the intrinsic extent of the nuclear \oiii component, we directly modelled the PSF and \oiii SW halo profile. To this end, we used the directly observed PSF (green data in Fig.\ref{fig:O3_sbp}), as measured in the central $0\farcs5$ to avoid the effect of the PSF wings. We convolved an unknown intrinsic \oiii surface brightness profile to ultimately reproduce the observed one (red data in Fig. \ref{fig:O3_sbp}). Assuming circular symmetry, the measured FWHM of the PSF is $0\farcs11$, in excellent agreement with the values found by \citet{deugenio2024fast}. Since we do not know a priori the intrinsic shape of the \oiii surface brightness profile in the nuclear region, we explored different possible distributions. We found the best agreement when using peaky (high excess kurtosis) distributions such as an exponential or a Lorentzian profile. In both cases we found an intrinsic FWHM of the \oiii profile of $0\farcs12$ (0.6 kpc).

\subsubsection{Kinematics of the \oiii emission}

Here we investigate the nature of the \oiii\ emission by studying the gas kinematics obtained from the R2700 data. The top-left inset of Fig.~\ref{fig:o3_map_spectra} shows the \oiii\ R2700 spectrum extracted in the central region around the QSO. This spectrum clearly reveals the presence of a broad blueshifted \oiii component. This component is shifted by --360 km $\rm s^{-1}$ with respect to the systemic redshift as traced by the BLR and has a FWHM of 1,030 km $\rm s^{-1}$. Based on the properties of the \oiii\ line profile, we expect that this component is associated with a galactic outflow.

\begin{figure*}[h!]
\centering
\includegraphics[width=\linewidth, trim=0mm 10mm 0mm 14mm,clip]{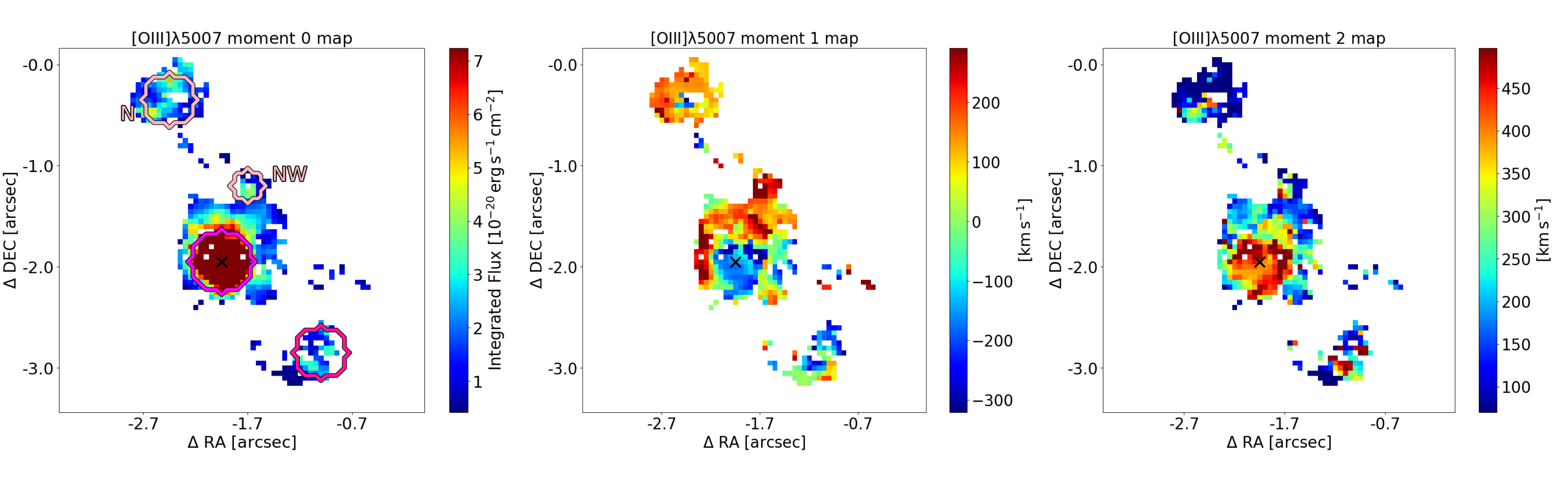}
\caption{Moment maps as derived from the best fit models of the high resolution spectra on each spaxel. The cross shows the QSO position. In the moment 0 map we also extraction regions for the nucleus and the \oiii halo as well as the other \oiii emitters.}
\label{fig:o3_mom_maps}
\end{figure*}

This interpretation is corroborated by the \oiii moment maps, shown in Fig.~\ref{fig:o3_mom_maps}. The moment maps were obtained by analysing the best-fit models between $\rm \pm 1,500 \, km \, s^{-1}$ from the expected \oiii location in each spaxel where this line was detected at $\rm \geq 2 \, \sigma$. Although the bulk of the \oiii emission is concentrated in the central region, the \oiii emitters as well as part of the extended emission are detected. The moment 1 map shows evidence for the presence of a ``V-shaped'' blueshifted \oiii component at the QSO location ascribable to the approaching side of an outflow (e.g. \citealt{carniani2015ionised, venturi2018magnum, mingozzi2019magnum, marconcini2023moka3d}). The high dispersion in the nuclear region observed in the moment 2 map provides evidence for the outflow nature of the \oiii emission at the QSO location. The \oiii linking ULAS J1342 to its NW neighbour clump appears to have low velocity dispersion, possibly indicating a low-turbulence exchange of gas between the two systems.

Whilst the outflow nature of \oiii emission at the centre is evident, the origin of the elongated component to the SW (see Fig.~\ref{fig:o3_map_spectra}) is not obvious, as it is not possible to perform a spaxel-by-spaxel kinematic analysis due to its faintness.  In Fig.~\ref{fig:o3_map_spectra} (top-right panel), we present the integrated high-resolution spectrum from a representative region of the extended \oiii\ emission, along with the corresponding spectral best-fit models. The low signal-to-noise ratio of the spectrum prevents us from determining whether one (dashed profile) or two Gaussian components (solid profile) are required to fit the line profile. Nonetheless, in both scenarios, the overall line FWHM exceeds 1,000 km $\rm s^{-1}$. This width is consistent with those observed in quasar-driven outflows at comparable bolometric luminosities and lower redshifts (e.g., $z \lesssim 1$, \citealt{wu2022catalog}; $z \sim 1.5$–$3.5$, \citealt{shen2016rest}; $z \gtrsim 3.5$, \citealt{perna2025ga, bertola2025ga}), lending support to the outflow interpretation for the extended emission.

\subsubsection{Outflow properties}

Under the assumption that both the nuclear and the extended \oiii are tracing the ionised component of an outflow, we can estimate key outflow properties such as the mass outflow rate ($\rm \dot{M}$) and the mass loading factor ($\rm \eta = \dot{M} /SFR$). Here we adopted the formalism described in \citet{carniani2015ionised} to estimate these outflow-related quantities. In particular, the ionised outflow mass ($\rm M_{out}$) can be written as:
\begin{equation}
    \rm M_{out} = 8 \times 10^7 \, \frac{K}{10^{[Z-Z_{\odot}]}} \, \left(\frac{L_{[OIII]}}{10^{44} \, erg \, s^{-1}}\right) \left(\frac{n_e}{500 \, cm^{-3}}\right)^{-1} M_{\odot}
\label{eq:Mout}
\end{equation}

Here, $\rm K = \langle n_e\rangle^2 / \langle n_e^2\rangle \sim 1$ is the electron density clumping factor, $\rm Z-Z_{\odot}$ is the logarithm of the gas metallicity with respect to the solar one, and $\rm n_e$ is the electron density. Several authors found that \oiii-based $\rm M_{out}$ systematically underestimate the total mass of ionised gas by a factor of $\sim$3 with respect to \Hb (e.g. \citealt{cano2012observational, carniani2015ionised, fiore2017agn, venturi2023complex, cresci2023bubbles}). We therefore increase the outflow mass by a factor of 3. To measure $\rm L_{[OIII]}$ for the nuclear component of the outflow, we used the luminosity of the outflow component (the blueshifted one in the spectrum in top-left panel in Fig. \ref{fig:o3_map_spectra}) from the high-resolution cube. As the nuclear spectrum was obtained by integrating on an aperture with a radius of $0\farcs3$, we employed the same aperture correction of 16\% described in Sect. \ref{sec:blr_fit}. We measured the \oiii luminosity of the extended component in the low-resolution spectrum obtained by summing, over the \oiii halo, all the spaxels where the \oiii was detected at $\geq 1\sigma$. Here we acknowledge that fluxes measured in the low- and high-resolution data for the same object differ by more than $\sim 10\%$ above $z=4.7$ (see e.g. \citealt{deugenio2025jades}). However, the systematic uncertainties due to the poor knowledge of the outflow electron density and geometry are far larger than such an effect.

Assuming a constant outflow velocity, the mass outflow rate is simply:
\begin{equation}
    \rm \dot{M}_{out} = C \,\frac{M_{out} v_{out}}{R_{out}}
\end{equation}

Where $\rm v_{[OIII], out}= v_{[OIII],off} + 2 \, \sigma_{[OIII]}$ (\citealt{rupke2013multiphase}), with $\rm v_{[OIII],off}$ being the absolute value of the offset velocity of the outflow component computed with respect to the systemic one and $\rm \sigma_{[OIII]}$ the velocity dispersion of the outflow \oiii component. Here $\rm R_{out}$ is the physical size of the region associated with the outflow. The coefficient C reflects the outflow history: $\rm C=1$ represents a constant outflow history, while $\rm C=3$ assumes a constant volume density in the outflow cone, implying a decaying outflow history (see Fig. 4 in \citealt{lutz2020molecular}).

For the nuclear component, we assumed the physical size of the outflow to be half the angular size of the intrinsic \oiii profile (see Sect.\ref{sec:extO3}), which is $0\farcs06$, corresponding to a projected size of $\sim$0.3 kpc. The outflow velocity was computed directly from the blueshifted component clearly visible in the nuclear spectrum (top left panel of Fig. \ref{fig:o3_map_spectra}), which is $\rm v_{out}$=--1220 $\rm km \, s^{-1}$. For the extended \oiii, we estimated $\rm R_{out}$ as the size of the \oiii halo, which is $1\farcs4$, corresponding to $\sim$7 kpc. Here, we used the outflow velocity derived from single Gaussian fit to the high-resolution spectrum (top-right panel in Fig. \ref{fig:o3_map_spectra}), which is consistent with the nuclear one. Assuming an average electronic density of $\rm n_e=200 \, cm^{-3}$ (e.g. \citealt{fiore2017agn}) and a gas metallicity of $\rm Z = 1.3 \, Z_{\odot}$ (\citealt{novak2019alma}), we derived a total mass outflow rate of $\sim$270 $\rm M_{\odot} \, yr^{-1}$, with the extended region  accounting for only $\sim$1\% of the total.

With the aim of exploring the possible systematics, we employed different values for both the outflow velocity and the electron density of the outflow. In particular, we also computed the outflow kinematics using the broad component of the high-resolution spectrum of the extended outflow (broad component in the top right panel of Fig.~\ref{fig:o3_map_spectra}), instead of that of the total single-component profile. We also changed the reference frame, computing the outflow kinematics with respect to the [\ion{C}{ii}] redshift as rest-frame (z=7.540, \citealt{venemans2020kiloparsec}) instead of the NLR one (z=7.535, Sect. \ref{sec:blr_fit}). These changes cause differences in $\rm \dot{M}$ of $\sim 40 \, \rm M_{\odot} \, yr^{-1}$ at most, which are negligible with respect to the uncertainty introduced by the unknown density of the ionised outflow. Because of this, we performed the calculation assuming both a high- and an intermediate-density scenario. In the first case we assumed a density $\rm n_e = 1000 \, cm^{-3}$, which translates into a lower limit for the mass outflow rate. Alternatively, we set $\rm n_e = 180 \, cm^{-3}$ as a lower limit, as derived from previous observations which found the \ion{H}{ii} regions in the ULAS J1342 environment to have $\rm n_e > 180 \, cm^{-3}$ (\citealt{novak2019alma}). This latter value sets, instead, an upper limit for the mass outflow rate. Under these extreme assumptions the mass outflow rate ranges between $\rm \sim 50-300 \, M_{\odot}$. Here, we note that we also did not correct for possible dust extinction, since we do not have indication for it. If present, extinction would reduce the observed \oiii flux, therefore the intrinsic (i.e. extinction-corrected) mass outflow rate would be higher than that estimated here.

The total ionised mass outflow rate is between $\rm \sim 50$--$300 \, M_{\odot} \, yr^{-1}$, depending on the values of $\rm n_e$ assumed, but it increases to $\rm \sim 150$--$900 \, M_{\odot} \, yr^{-1}$ in the case of $\rm C=3$. Because of the unknown outflow history, such factor should be folded into the systematic uncertainty. This range overlaps, for a good part, with the star formation rate ($\rm SFR$), which, according to different indicators ([\ion{C}{ii}]; \citealt{venemans2017copious}, FIR luminosity; \citealt{novak2019alma, venemans2020kiloparsec}), lies in the range 85--545 $\rm M_{\odot} \, yr^{-1}$. As a consequence, the mass loading factor ($\rm \eta = \dot{M}/SFR$) spans a wide range of values between 0.1 and 10. It is therefore possible that even the $\rm \dot{M}_{out}$ carried by the ionised component of the on-going outflow alone could surpass the star formation in ULAS J1342. However, tighter constrains on the electron density of the outflow and its geometry are needed to assess this possibility. Nonetheless, we must also recall that the $\rm \dot{M}_{out}$ estimated here is likely a lower limit to the total mass outflow rate. Outflows prompted by luminous quasars are indeed expected to stratify on multiple phases, with the neutral atomic and molecular ones carrying the largest shares of mass and energy (e.g. \citealt{baron2019discovering, fluetsch2019cold,veilleux2020cool, fluetsch2021properties, davies2024jwst, deugenio2024fast}). For this reason, we point out that the $\rm \dot{M}_{out}$ estimated here only represents a lower limit to the total one, and it is possible that its entirety could easily exceed the host galaxy SFR, therefore delivering an early feedback. Yet, no strong outflows have been so-far detected in [\ion{C}{ii}] observations of ULAS J1342 (\citealt{banados2019z}).

\begin{figure}[h!]
\centering
\includegraphics[width=\linewidth,clip]{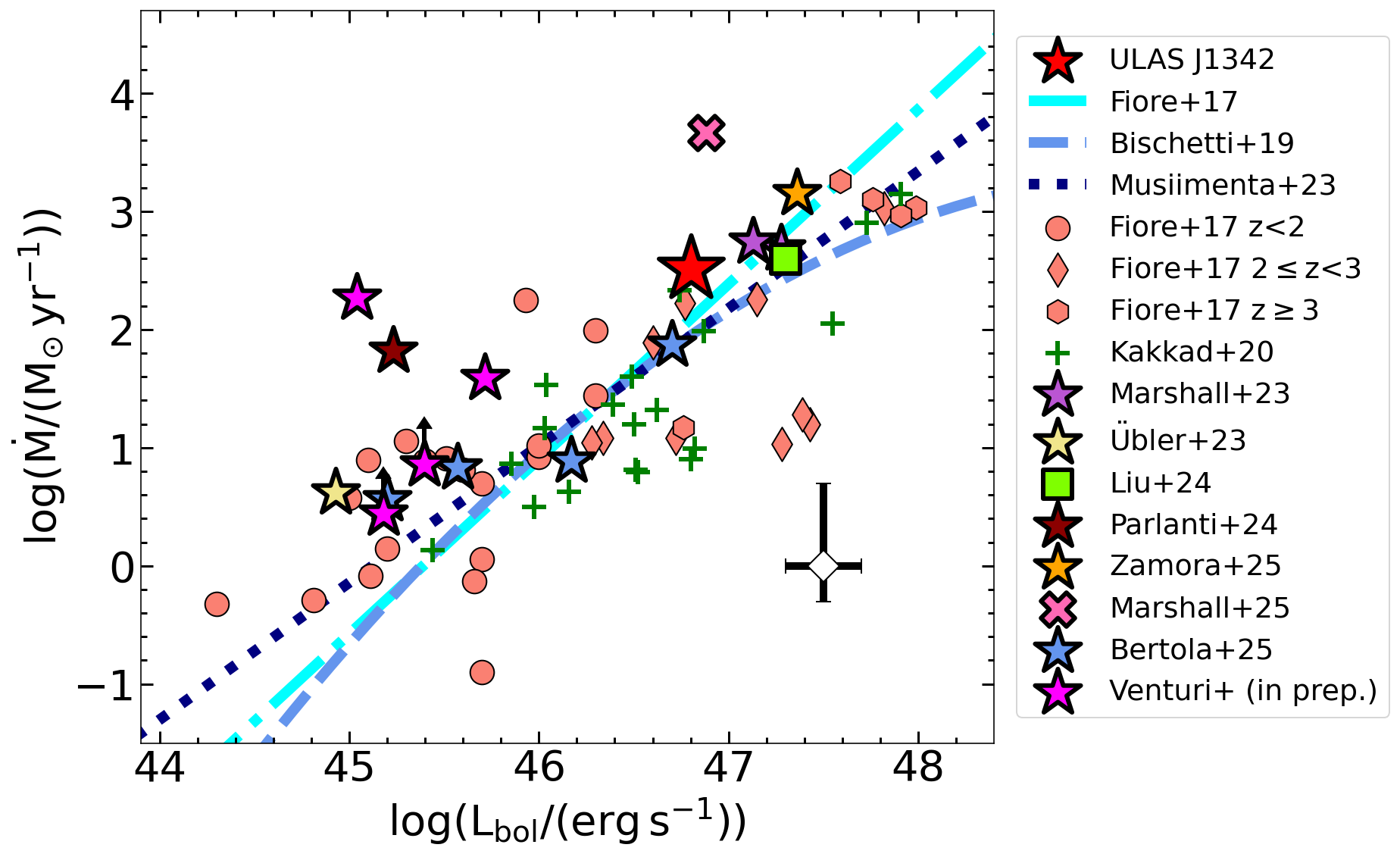}
\caption{Mass outflow rate against bolometric luminosity of the quasar (red star), in the context of other literature sources (\citealp{fiore2017agn}, \citealp{kakkad2020super}, \citealp{marshall2023ga}, \citealp{liu2024fast}, \citealp{marshall2025jwst}). Sources from F17 have been divided according to their redshift. Thick-edged sources have been observed with JWST NIRSpec IFU, while stars mark targets within the GA-NIFS programme. The best-fit lines have been scaled to match the mean value of the distribution (excluding ULAS J1342). The vertical errorbars on the hollow diamond in the bottom right corner exemplify of the spread introduced by changing the density from 100 $\rm cm^{-3}$ to 1000 $\rm cm^{-3}$. The horizontal errorbar shows a typical uncertainty on the bolometric luminosity of 0.2 dex.}

\label{fig:Lbol_Mdot}
\end{figure}

In Fig. \ref{fig:Lbol_Mdot} we compare the outflow rate measured in ULAS J1342 with that of other type 1 AGN and QSOs at different bolometric luminosities and redshifts where the outflow properties have been determined based on IFU data. In particular, the values presented there have been homogenised according to the same procedure described in Venturi et al. in prep. Here we outline the main features. All the outflow velocities ($\rm v_{out}$) were recomputed adopting the definition $\rm v_{out}= v_{off} + 2 \, \sigma$ with $\rm v_{off}$ being the velocity difference between the outflow and the rest component and $\rm \sigma$ the velocity dispersion of the line outflow component. The outflow radius was computed as the maximum radius reached by the outflow as traced by the broad component. The ionised outflow mass was computed using Eq.\ref{eq:Mout} in case the outflow was detected in the \oiii line or the equivalent expression for the \ion{H}{$\alpha$} derived in \citet{carniani2015ionised}. Lastly, the geometrical factor was set to $\rm C$=3 to comply with other outflow compilations assuming the case of constant average volume density. Furthermore, for all the samples shown in Fig. \ref{fig:Lbol_Mdot} we used a constant electronic density $\rm n_e=500 \, cm^{-3}$. Changing this value for all the sources would simply result in a vertical shift of the data points in the plot.

The mass outflow rate detected in ULAS J1342 nicely aligns with the main trend derived from the other samples, as well as with the empirical relations derived by \citet{fiore2017agn}, \citet{bischetti2019gentle} and \citet{musiimenta2023new}, shown in Fig.~\ref{fig:Lbol_Mdot} as dot-dashed, dashed, and dotted line, respectively. Interestingly, as already noted in \citet{bertola2025ga} for AGN at $z\gtrsim 1$, the relation between mass outflow rate and luminosity does not seem to exhibit changes with the redshift. High-$z$ QSOs sit on the trends observed at lower redshifts and similar luminosities within the large observed scatter. Alternatively, it is also possible that any redshift evolution could be diluted by the large systematic uncertainties, for instance on the outflow electron density, which might be evolving with redshift (e.g. \citealt{isobe2023redshift, topping2025aurora}).

Lastly, the spatial extent of the \oiii revealed by the low-resolution data suggests an elongated structure towards the SW direction. This component, morphologically separated from a nuclear outflow, could be related to a past outflow event which has propagated through the host galaxy. Within this picture, which we highlight to be mostly speculative, and assuming a roughly constant outflow velocity of 1,200 $\rm km \, s^{-1}$, the outflow should have been launched $\sim 5.7$ Myr earlier, to propagate on a projected scale of $\sim$7 kpc (see Fig. \ref{fig:O3_sbp}). In order to test this hypothesis, deeper high spectral resolution observations will be needed.

\subsection{Black hole mass and Eddington ratio}
\label{sec:accretion_parameters}

We used the best fit parameters from the spectral analysis to derive the accretion parameters ($\rm M_{BH}$, $\rm L_{bol}$) using standard optical bolometric corrections ($\rm k_{bol}$) for $\rm L_{bol}$ and the SE calibrations for $\rm M_{BH}$.

In particular, we adopted the widely employed \citealp[(VP06)]{vestergaard2006determining} and the \citealp[(S11)]{shen2011} calibrations for the \Hb and the \ion{Mg}{ii} respectively, using our measured \Hb and \ion{Mg}{ii} broad line widths and continuum fluxes measured from the high and low resolution spectra. With the aim of exploring also SE calibrations which take into account the possible effect of different Eddington ratios on the BLR shape, we also employed the recent \citealp[(P25)]{pan2025iron} calibration. This new recipe takes advantage of the $R-L$ relation derived in \citet{du2019radius}, which uses the \ion{Fe}{ii}\textsubscript{opt}/\Hb ratio (see also Sect. \ref{sec:BLR_chem_enr}) to account for the effect of the accretion rate on the BLR geometry.

We estimated the bolometric luminosity using the 3,000 \AA\, monochromatic luminosity and assuming an empirical fixed bolometric correction $\rm k_{bol}=5.15$ from \citealp[(R06)]{richards2006spectral} as well as an empirical and a theoretical luminosity-dependent correction respectively by \citealp[(R12)]{runnoe2012updating} and by \citealp[(N19)]{netzer2019bolometric}\footnote{Here we highlight that the N19 bolometric corrections have been computed assuming an average inclination of the LoS with respect to the disc axis of 57$^{\circ}$. These corrections should be smaller (by a factor of $\sim$1.5) in the case of an inclination of 30$^{\circ}$, which is more suitable for broad line quasars where the dusty torus avoids nearly equatorial LoSs. Such a correction would enhance the agreement with our model \lbol.}. We report all the values of $\rm M_{BH}$ and $\rm L_{bol}$ in Table \ref{tbl:accretion_pars}. In all these measurements the systematic uncertainties are far larger than the statistical ones.

The \mbh\, estimated using the emission line calibrations for \Hb and \ion{Mg}{ii} are respectively $\rm \log(M_{BH}/(M_{\odot}))$ = 9.0 (VP06) and 9.3 (S11) with a systematic uncertainty of $\sim$0.5 dex. The \mbh~ derived employing the P25 prescription is smaller, $\rm \log(M_{BH}/(M_{\odot}))$ = 8.6, as a consequence of the smaller radius of the BLR assumed by their prescription, with a similar systematic uncertainty. The \lbol\, derived using the aforementioned bolometric corrections are respectively $\rm \log(L_{bol}/(erg \, s^{-1}))=$ 47.1 (R06) and 46.8 and 46.9 respectively for R12 and N19 with systematic uncertainties of the order of $\sim$0.2 dex. We compare the values derived from these calibrations with those coming from the accretion disc modelling in the bottom left panel of Fig. \ref{fig:ad_fit} and in Table \ref{tbl:accretion_pars}. By combining these estimates, the Eddington ratio of ULAS J1342 varies from a minimum of $\rm \lambda_{Edd}\sim$0.3 to a maximum of $\rm \lambda_{Edd}\sim$2.8 with systematic uncertainties between 0.4--0.5 dex.

For the accretion disc modelling, although the innermost radius (or equivalently $a$) and the inclination are very loosely constrained, all of the $\rm M_{BH}-L_{bol}$ joint posterior distributions from the individual models show clear uni-modal maxima. The best-fit values of the three models are consistent within the $1\sigma$ uncertainties and yield $\rm \log(M_{BH}/M_{\odot})=9.2$ and $\rm \log (L_{bol}/(erg \, s^{-1}))=46.8$. The uncertainties are of the order of 0.2 dex and 0.1 dex for $\rm M_{BH}$ and $\rm L_{bol}$ respectively, of the same magnitude as those derived employing the ``BADFit'' routine and in line with other works employing accretion disc modelling to estimate the accretion parameters (e.g. \citealt{lai2023characterising, wolf2024accretion}). 

The disc modelling estimates are also consistent with the \Hb and \ion{Mg}{ii} estimates, yet this is mostly due to the large uncertainties on single-epoch virial masses (0.4--0.5 dex). The bolometric luminosity is instead consistent with the N19, but not with the R06. This is, however expected, since the fixed R06 $\rm k_{bol}$ overestimates the total luminosity in the case of luminous sources (see e.g. Fig. 2 in N19). Our fiducial estimate of the Eddington ratio of ULAS J1342 from accretion disc modelling is $\rm \lambda_{Edd} \sim 0.4$. Therefore the disc is in a state of moderate accretion, and could be outside of the geometrically thin regime, yet it does not appear to exceed the Eddington limit.

\begin{table}[h!]
\centering
\setlength{\tabcolsep}{3pt}
\begin{tabular}{c c c c}
 \hline \noalign{\smallskip}
 $\rm M_{BH}$ calibration & $\rm \log(M_{BH}) $ & $\rm L_{bol}$ calibration &  $\rm \log(L_{bol})$ \\
        & $\rm M_{\odot}$     &        & $\rm erg \, s^{-1}$  \\

 \hline \noalign{\smallskip}
 $\rm H\beta$ VP06   &  $9.0 \pm 0.5$  & R06            & $47.1 \pm 0.1$ \\ 
 $\rm H\beta$ P25    &  $8.6 \pm 0.5$  & R12            & $46.8 \pm 0.2$ \\ 
 $\rm MgII$   S11    &  $9.3 \pm 0.5$  & N19            & $46.9 \pm 0.2$ \\ 
 This work (AD)      &  $9.2 \pm 0.2$  & This work (AD) & $46.8 \pm 0.1$ \\ 

 \hline
\end{tabular}

\caption{Accretion parameters estimated using either empirical calibrations (see Sect. \ref{sec:accretion_parameters}) or accretion disc modelling (AD) which are used hereafter as fiducial values. The uncertainties on the calibrated black hole masses are representative of the typical ones for the proposed emission line. The same holds for the bolometric luminosities derived from monochromatic luminosities.}
\label{tbl:accretion_pars}
\end{table}

\subsection{The growth of the ULAS J1342 SMBH}

Under general assumptions, it is possible to attempt inferring the evolutionary pathway ULAS J1342 went through. Previous estimates showed that, assuming an Eddington limited accretion ($\rm \lambda_{Edd}=1$), a starting seed of $\rm 10^5 \, M_{\odot}$ is required at $z\sim15$ (see Fig. 5 in \citealt{campitiello2019black}). Since we infer a $\rm \lambda_{Edd} \sim 0.4$, the highly accreting phase must already be over by $\rm z\sim7.5$ with ULAS J1342 having assembled most of its mass within the first 0.5 Gyr of cosmic time. The current accretion rate appears too low even for `heavy seeds' to produce the observed mass. This is in line with SMBH accretion models which predict a super-Eddington phase onto the BH seed likely in a gas enshrouded environment, concealing the intrinsic quasar SED, followed by a steep transition to sub-Eddington accretion in a non-obscured phase (\citealt{lapi2014coevolution}).

\begin{figure}[h!]
\centering
\includegraphics[width=\linewidth,clip]{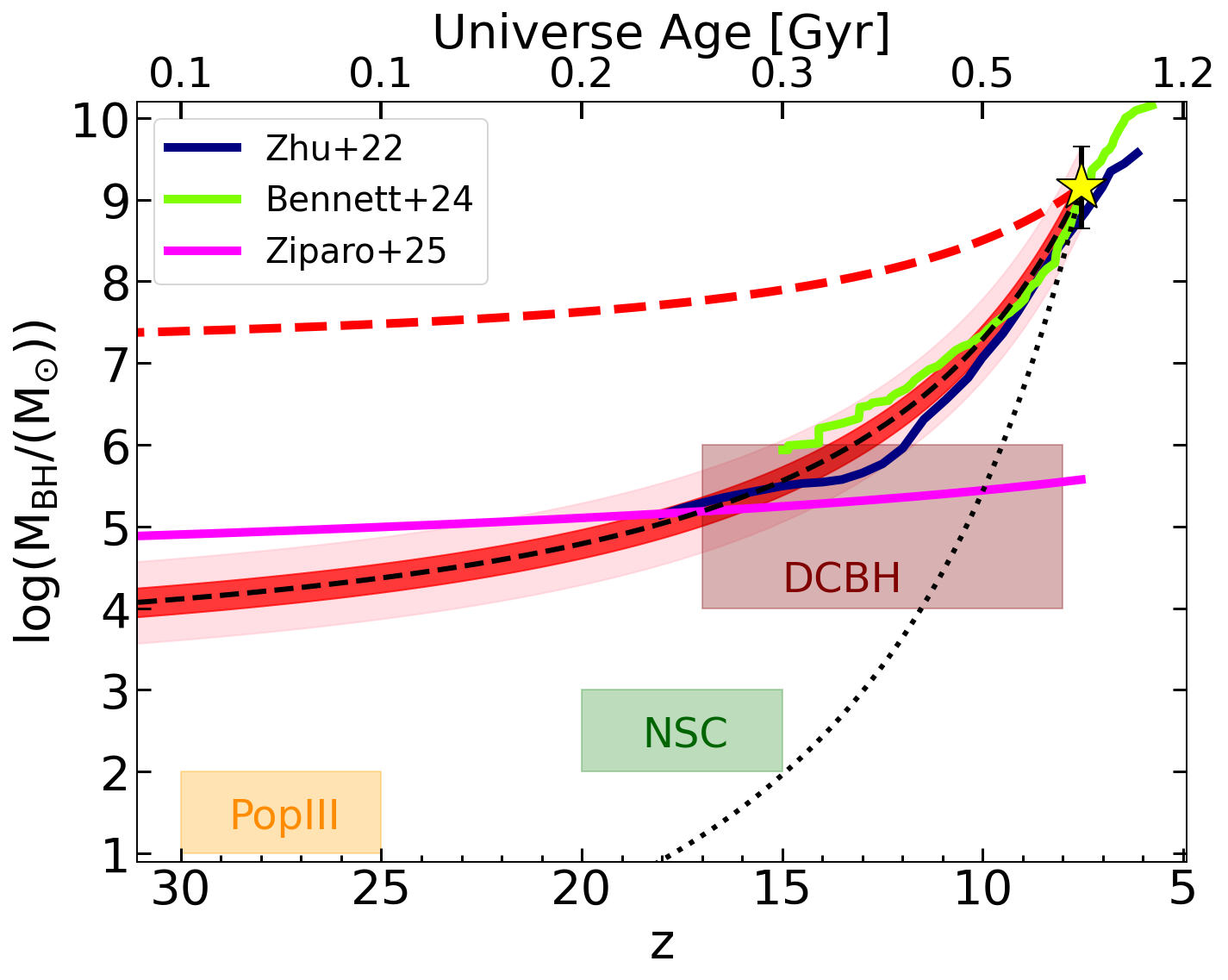}
\caption{Possible accretion histories of the SMBH hosted in ULAS J1342. The red dashed line represents the simple exponential accretion history with unitary duty cycle and the observed $\rm \lambda_{Edd}=0.4$, while the black dashed and dotted ones respectively assume $\rm \lambda_{Edd}=1$ and 2. The red (pink) stripe highlights the region around $\rm \lambda_{Edd}=1$ encompassed by the best fit values using the black hole mass as estimated from the accretion disc modelling (SE). Simulated accretion histories for massive halos by \citet{zhu2022formation} and \citet{bennett2024growth} are shown in blue and green respectively. Rectangular patches show the regions of this parameter space populated by black hole seeds according to different seeding mechanisms (see text).} 
\label{fig:Mbh_z}
\end{figure}

In Fig. \ref{fig:Mbh_z} we show possible accretion histories aimed at reproducing the observed SMBH in ULAS J1342. There, we assumed a fixed average Eddington ratio (our current estimate of $\rm \lambda_{Edd}=0.4$ red dashed, $\rm \lambda_{Edd}=1$ black dashed line, $\rm \lambda_{Edd}=2$ black dotted line) and a unitary duty-cycle. We also show the accretion histories derived for black holes hosted in massive galaxies in recent simulations resulting in SMBH masses akin to ULAS J1342. For their simulation framework, \citet{zhu2022formation} (blue line in Fig.~\ref{fig:Mbh_z}) targeted the largest halos in 15 simulations up to redshift 6 following the evolution of the SMBH and adopting different prescriptions for a set of seeding, accretion and feedback parameters (we refer to their fiducial S5-REF set of parameters, see their Table 1). Instead, \citet{bennett2024growth} (green line) employed the \textsc{fable} suite of simulations (\citealt{henden2018fable}) with a modified setup tuned to find a plausible pathway to grow massive black holes by $\rm z \sim 6$ (the `Reference' simulation, see their Sect. 2.3). These evolution histories encompass our simple exponential $\rm \lambda_{Edd}=1$ accretion, and manage to reproduce the observed mass of ULAS J1342. 

The boxes in Fig. \ref{fig:Mbh_z} highlight possible ranges of seed black hole masses according to different seeding scenarios. In particular, we show the parameter spaces occupied by: i) light seeds originating from remnants of massive metal-free stars inhabiting early mini halos between $z\sim 20$--$30$ (PopIII, $\rm M\sim10^{1}$--$10^{2} \, M_{\odot}$, e.g. \citealt{madau1998star, heger2003massive, yoshida2006formation}), ii) intermediate seeds  between $z\sim 10$--$20$ ($\rm M\sim10^{2}$--$10^{3} \, M_{\odot}$) produced by nuclear star clusters (NSC) in primordial galaxies (e.g. \citealt{devecchi2009formation, davies2007, lupi2014constraining}), iii) heavy seeds where black holes formed via atomic cooling direct collapse in halos containing pristine gas at $z\sim 8$--$17$ (DCBH; e.g. \citealt{eisenstein1994origin, silk1998quasars, ferrara2014initial}).
The redshift ranges highlighted here reflect those where the formation channels are most likely to produce the respective seeds. A later seed formation for black holes originating from PopIII stars or NSCs would necessitate an even more extreme accretion history. We also show the expected mass evolution for PBHs predicted by \citet{ziparo2025primordial} (see their Eq. 3.2) which, appearing at $z>20$, have assembled by $z\sim$10 a mass in the range of `heavy seeds' (\mbh$\gtrsim \, 10^{5} \, M_{\odot}$). A phase of moderate, yet still sub-Eddington, accretion of a PBH could also explain the observed mass in ULAS J1342.

Unless the accretion history of ULAS J1342 significantly differs from those predicted by the simulations for massive halos, or alternatively deviates from them with a decisively super-Eddington phase at $\rm z \gtrsim 15-20$, heavy seeds seem to be the preferred scenario to produce a mass of about $\rm 10^9 \, M_{\odot}$ at $z\sim7.5$.

\subsection{BLR chemical enrichment}
\label{sec:BLR_chem_enr}
\begin{figure}[h!]
\centering
\includegraphics[width=\linewidth,clip]{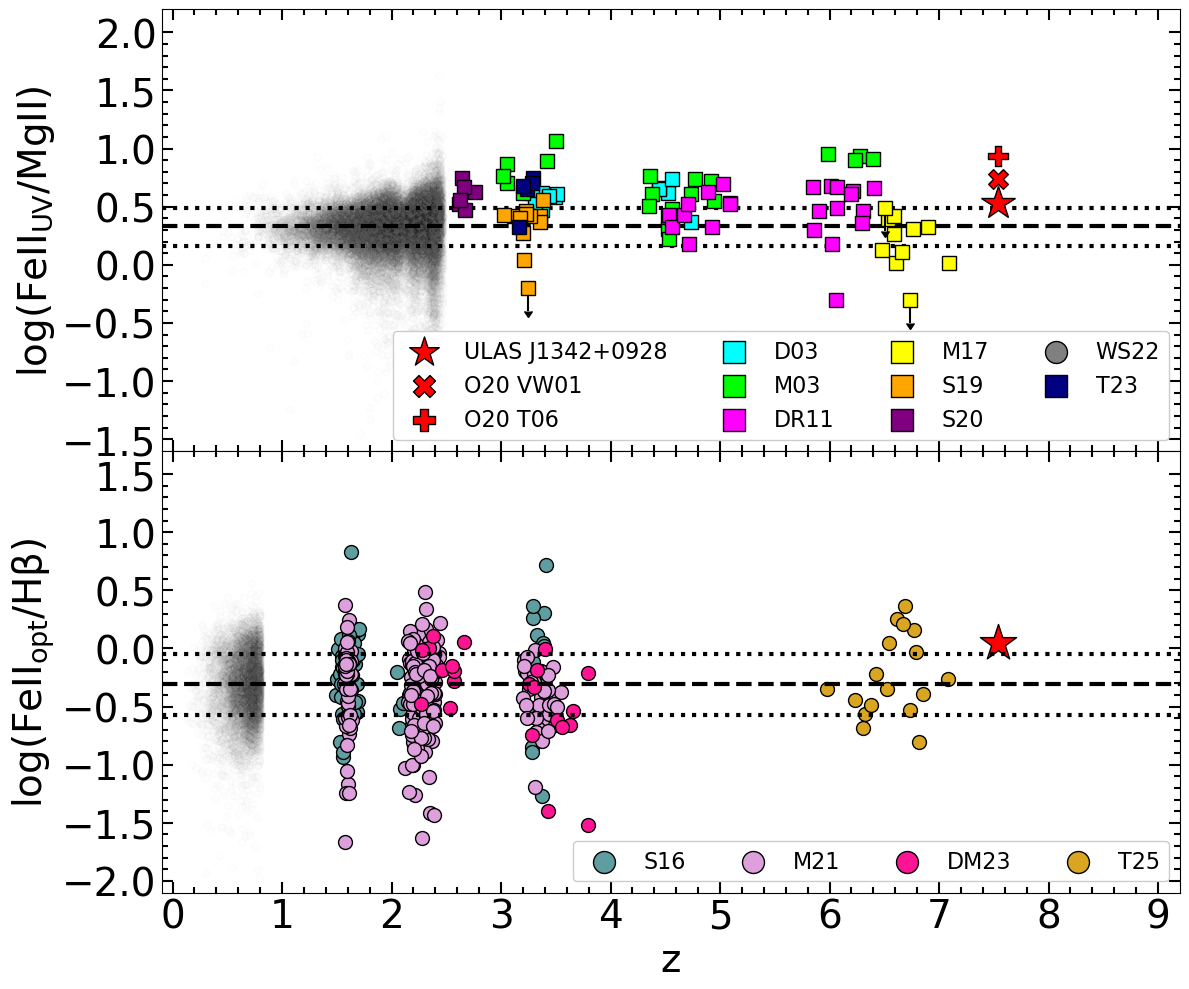}
\caption{The \FeMg and \ion{Fe}{ii}/\Hb ratios of ULAS J1342 (red star) in the context of the redshift evolution of such ratios, together with other literature samples. The red "$\rm \times$ and $+$" mark the values already estimated by \citealp[(O20)]{onoue2020no} for ULAS J1342 adopting the \citealp[(VW01)]{vestergaard2001empirical} and the \citealp[T06]{tsuzuki2006fe} \ion{Fe}{ii}\textsubscript{UV} templates respectively and a Gaussian line profile for \ion{Mg}{ii}. The dashed line marks the median value of the low redshift SDSS sample from WS22.}
\label{fig:RFe_z}
\end{figure}

The metal abundance in quasar BLRs allows us to sample the metallicity of gas up to remote cosmic epochs. Photoionisation calculations suggest that ratios of broad UV lines (e.g. (\ion{Si}{iv}+\ion{O}{iv})/\ion{C}{iv}, (\ion{C}{iii}]+\ion{Si}{iii}])/\ion{C}{iv}, \ion{Al}{iii}/C IV, \ion{N}{v}/\ion{C}{iv}) are suitable to estimate the BLR metallicity (e.g.
\citealt{hamann2002metallicities,nagao2006}). Several works demonstrated the feasibility of this approach up to z$\sim$7 (see e.g. \citealt{wang2022metallicity} and references therein). However, most of these broad UV lines are often blended and fall in the NIR band above z$\sim$4. Because of this, the ratio between \ion{Fe}{ii}\textsubscript{UV} and \ion{Mg}{ii} has been widely employed as a first order proxy of the Fe/$\alpha$ element abundance ratio (e.g. \citealt{dietrich2003fe, jiang2007gemini, derosa2011evidence, mazzucchelli2017physical, shin2019fe, trefoloni2023most}). No clear redshift evolution for this ratio was found up to z$\sim$7. At the same time, it has also been pointed out that the dependence of such a ratio on the actual BLR metallicity is rather weak (\citealt{sarkar2021improved}). A more suitable proxy for the actual BLR metallicity is instead the \ion{Fe}{ii}\textsubscript{opt}/\Hb (e.g. \citealt{trefoloni2024missing}), which however falls deep into the NIR band for this redshift range. Because of this, as we show in Fig.~\ref{fig:RFe_z}, all of the \ion{Fe}{ii}\textsubscript{opt}/\Hb measurements in the pre-\jwst era were limited to $z<4$.

In Fig. \ref{fig:RFe_z} we show both the \ion{Fe}{ii}\textsubscript{UV}/\ion{Mg}{ii} and the \ion{Fe}{ii}\textsubscript{opt}/\Hb ratios evaluated for the nuclear region of ULAS J1342, together with other quasar samples at different redshifts (\citealp[D03]{dietrich2003fe}, \citealp[M03]{maiolino2003early}, \citealp[DR11]{derosa2011evidence}, \citealp[S16]{shen2016rest}, \citealp[M21]{matthews2021placing}, \citealp[M17]{mazzucchelli2017physical}, \citealp[S19]{shin2019fe}, \citealp[S20]{sameshima2020mg},  \citealp[WS22]{wu2022catalog}, \citealp[T23]{trefoloni2023most}, \citealp[DM23]{deconto2023high}, \citealp[T25]{trefoloni2024missing}). The EW of UV and optical \ion{Fe}{ii} have been computed by integrating the best fit models over the wavelength ranges 2,200--3,090\AA\, and 4,434--4,684\AA\, respectively. In order to roughly match the ULAS J1342 luminosity regime, we only selected sources above $\rm \log (L_{3,000\AA}/(erg \, s^{-1}))=44.0$ for the \ion{Fe}{ii}\textsubscript{UV}/\ion{Mg}{ii} panel. For the \ion{Fe}{ii}\textsubscript{opt}/\Hb panel we applied instead a selection in \Hb luminosity, by requiring $\rm \log (L_{H\beta}/(erg \, s^{-1}))\geq 43.0$. Despite being the highest redshift measurements of these ratios, our estimates do not show any hint of decrease with respect to lower redshift samples. Nonetheless, we caution that several systematic effects, that we discuss in Appendix \ref{app:fe2mg2_systematics}, could dilute a possible weak evolution. Both the broad line ratios measured in ULAS J1342 are in line with the values measured above $z\sim2$ and on the upper envelope of the SDSS distribution (the median and 16$^{\rm th}$--84$^{\rm th}$ percentiles are shown respectively as a dashed and dotted lines), although some systematics regarding the fitting technique and the \ion{Fe}{ii} templates hamper a homogeneous comparison among these samples. If these ratios are actually tracing the BLR metallicity, albeit in conjunction with other parameters, these measurements testify its lack of evolution up to $z\sim7.5$, at least for the brightest QSOs.

\section{Discussion}
\label{sec:discussion}

Despite being already one of the most well-studied high-redshift QSOs, deep into the epoch of reionisation, the new \jwst observations, obtained within the GA-NIFS programme, revealed several interesting details about ULAS J1342.

The optical spectrum covering the \Hb--\oiii region revealed an extremely complex \Feii emission, with relatively narrow multiplet peaks, closely resembling those observed in local Seyfert 1 galaxies (e.g. I Zwicky 1; \citealt{phillips1976optical, veron2004unusual}). A faithful modelling of such emission proved key to reliably subtract the unresolved BLR emission on a spaxel-by-spaxel basis.

The detection of spatially resolved \oiii emission on kpc scales follows those already reported in other high-$z$ QSOs observed by \jwst, such as DELS J0411-0907 and VDES J0020-3653 (\citealt{marshall2023ga}), J1007+2115 (\citealt{liu2024fast}) or NDWFS J1425+3254 (\citealt{marshall2025jwst}) at redshifts between $\rm z\sim 6.8$--$7.5$. Here, we detected two \oiii clumpy structures (the \oiii emitters) as well as a contiguous low-surface brightness region. One of the \oiii emitters is close ($\sim$4 projected kpc) to the QSO, to which it appears to be linked by an \oiii bridge, confirming the on-going interaction between these two objects, similarly to the system described in \citet{marshall2025jwst}. The other one appears as a separated system at $\sim$10 projected kpc, yet still at the same redshift as the QSO. 
Regarding the extended \oiii region, the most likely explanation for this component is a past outflow episode. Although the bulk of the outflow mass is concentrated within the central kpc, the diffuse component extends for $\sim$7 kpc, well above the typical galaxy size at these redshifts. For instance, assuming the typical galaxy size at high redshift to be described by the log-linear relation estimated by \citet{morishita2024enhanced}, we find a UV half-light radius $\rm R\sim 1 \, kpc$. Based on such a size, we expect this outflow to have already escaped the host environment injecting energy and momentum in the CGM as predicted by cosmological simulations (e.g. \citealt{costa2014environment, costa2018driving}). As already discussed in \citet{liu2024fast} for J1007+2115, a quasar similar to our target both in terms of black hole mass and redshift, the outflow velocities appear sufficient to exceed the escape velocity. Hydrodynamical simulations for analogous black hole masses predict the escape velocity from the halo to be $\rm \lesssim 1,000 \, km \, s^{-1}$ below $\rm \sim 1~ kpc$ (\citealt{ni2018connecting}), and the outflow velocities computed for ULAS J1342 make it slightly able to escape. Outflows piercing through the IGM are also expected to clear the path for the quasar radiation to escape the galaxy promoting the ionisation of the surrounding medium and the formation of \ion{Ly}{$\alpha$} nebulae both predicted (e.g. \citealt{costa2022agn}) and observed around quasars (e.g. \citealt{farina2019requiem}).

Combined information from the high- and low-spectral resolution data on the nucleus, allowed the black hole mass and luminosity of the QSO hosted in ULAS J1342 to be robustly constrained. In addition to the black hole masses estimated via the usual virial calibrations, we employed a SED-fitting approach to measure $\rm M_{BH}$ taking advantage of different accretion disc models. The result is consistent with the virial estimations, also due to the large systematics rooted in the SE estimates. Interestingly, this approach had already been attempted on ULAS J1342 by \citet{campitiello2019black}, who used accretion disc models to fit the Magellan/FIRE and Gemini/GNIRS spectra, covering roughly between rest-frame 1,200--3,000 \AA, presented in \citet{banados2018}. There, the authors obtained black hole masses in the range $\rm \log(M_{BH}/ M_{\odot})\sim 8.9$--$9.6$, in broad agreement with our findings.

In a broader context, our approach to the mass estimate problem adopting an accretion disc modelling also proved the capabilities of employing NIRSpec/PRISM observations. Accretion disc modelling has several key requirements, mostly a simultaneous (i.e. not taken at different epochs) large wavelength coverage, with a resolution still high enough to isolate continuum windows. Because of these requirements, this approach has generally been tested on relatively small samples with VLT/XSHOOTER (see e.g. \citealt{capellupo2015active,  lai2023characterising, wolf2024accretion}) or local samples with an exquisite coverage (e.g. \citealt{campitiello2020estimating}). Additionally, the peak of the accretion disc should be covered by the data. Observations only covering the $\rm \nu L_{\nu}\sim \nu^{1/3}$ region of the disc SED (generally falling in the near-UV/optical) would only provide upper limits to the black hole mass. This technique is also prone to systematic effects, such as host galaxy contamination as well as intrinsic reddening. However such effects are generally minimised in the case of luminous blue quasars like ULAS J1342. Our new low-resolution data fulfilled all these requirements, ultimately allowing us to derive an accretion disc-based $\rm M_{BH}$ that is more tightly constrained than those coming from single-epoch calibrations ($\sim0.2$ dex against 0.4--0.5 dex, see e.g. \citealt{shen2013mass} and references therein).

We explored different accretion disc models, namely a custom made \citet{shakura1973black} model, a version including relativistic effects ({\tt KERRBB}; \citealt{li2005multitemperature}), and also one suited for high accretion rates where the vertical radiative energy transport is not negligible ({\tt SLIMBH}; \citealt{skadowski2011relativistic}). The exact details of the disc structure are not currently fully understood, and accretion disc theory fails to explain several observables such as the variability timescales (see e.g. \citealt{lawrence2018quasar}), and the size (\citealt{morgan2010quasar, fausnaugh2016space,jiang2017detection}), while also the stability of the disc and the details behind the viscosity are questioned (see \citealt{abramowicz2013foundations} for a comprehensive review). However, quasar accretion discs must be efficient ($\rm \eta \sim 0.1$, e.g. \citealt{yu2002observational, shankar2008self}) in order to produce the enormous amounts of light observed while still abiding by the constraints given by the distribution of relic SMBH masses, i.e. the So{\l}tan argument (\citealt{soltan1982}). Additionally, the temperature profile, ultimately responsible for the steady-state SED, depends on the mass and the accretion rate, but not on the details of the mechanism providing the viscosity.
This consideration explains why the best fit values for both luminosity and black hole mass are extremely close ($\rm \lesssim 0.1$ dex) for all the explored models: to reach the observed luminosity, an efficient accretion process is required, and such feature underlies all these models\footnote{For slim accretion discs the efficiency itself depends on the accretion rate, however in the range of reasonable accretion rates for quasars (i.e. $\rm 0.01 \lesssim \lambda_{Edd} \lesssim 10$), the accretion process remains efficient (see e.g. Fig.~9 in \citealt{skadowski2009slim}). We also explored if an accretion rate dependent efficiency could change the accretion parameters derived here adopting, for instance the prescriptions in \citet{abramowicz2010leaving}, but found negligible differences.}.

Interestingly, this approach provides evidence for a moderate accretion state for this source ($\rm \lambda_{Edd}\sim 0.4$). Albeit not strictly super-Eddington, at such accretion rate the disc is likely thicker than classical thin discs, and the conditions for the launch of powerful nuclear outflows (\citealt{zubovas2013bal, nardini2015black, king2015powerful, nardini2019towards}) could be in place. These strong outflows are then expected to propagate into the host galaxy, ultimately delivering the feedback (e.g. \citealt{sijacki2007unified, dimatteo2008direct, harrison2018agn}). Such an effect must be in place at early times in order to justify the presence of already quiescent galaxies at $\rm z\gtrsim 3$ (e.g. \citealt{santini2021emergence, carnall2023massive, deugenio2024fast, russell2024cosmic}). Evidence is being collected for strong \oiii outflows, clear signatures of this mechanism, being at work already at $\rm z\sim 6$ (e.g. \citealt{marshall2023ga, yang2023spectroscopic, liu2024fast, loiacono2024quasar}). The ionised mass outflow rate detected in ULAS J1342 could in theory exceed the SFR, and ultimately deliver significant feedback to its host. There are, however, large uncertainties on both the SFR and the outflow density and geometry which prevent us from drawing more certain conclusions. We also remind that a large amount of gas could still be expelled in other phases (cold neutral, molecular), and therefore the total expelled gas could easily reach, or even exceed, a mass loading factor of unity.

From a chemical enrichment standpoint, it is interesting to note that the broad line ratios characterising the BLR of ULAS J1342 do not appear to deviate from those observed at lower redshift, as already pointed out in \citet{onoue2020no}. Broad metal line ratios such as \ion{Si}{ii}/\ion{C}{iv}, \ion{Si}{iv}/\ion{C}{iv}, \ion{Al}{iii}/\ion{C}{iv}, often interpreted as proxies to the BLR metallicity (see e.g. \citealt{lai2022chemical} and references therein), had already been investigated and compared to other lower redshift samples ($z<4$) broadly finding a lack of evolution. The \jwst observations presented here, in the rest-UV spectral region, previously unexplored at these high redshifts, highlighted how also the \ion{Fe}{ii}\textsubscript{opt}/\Hb ratio, regarded as a proxy to the BLR metallicity, follows the trends derived at lower redshift. The seeming non-evolution of \ion{Fe}{ii}\textsubscript{opt}/\Hb and \ion{Fe}{ii}\textsubscript{UV}/\ion{Mg}{ii} line ratios complies with the same findings for other broad line ratios (e.g. \citealt{nagao2006}). Notably, other properties characterising the broad lines of quasars, such as the EW (e.g. \citealt{croom2002correlation, stepney2023no}) have also been argued not to evolve with redshift. Putting together these clues, there is growing evidence for an early assembly and chemical enrichment of the BLR in quasars. In particular, as the \FeHb traces the enrichment of iron, produced by type Ia supernovae, it is puzzling to find a Fe emission at $z=7.54$ similar to those observed locally. This requires an early onset of the star formation likely around the epoch of the formation of the luminous galaxies at very high-z or, alternatively, conspicuous Fe production by alternative channels such as pair-instability supernovae.

An early chemical enrichment in QSOs is consistent with these sources residing in overdensities of the cosmic web (\citealt{cantalupo2014, farina2017mapping, balmaverde2017primordial, ota2018large, garcia2017mid}), which likely trace the most advanced stage of the galactic evolution at every redshift. This also provides a key justification for the luminosity--metallicity relation ($\rm L$--$Z$ relation; \citealt{hamann1993chemical, dietrich2003quasar}). Alternatively, the parameters ultimately defining the cloud properties (e.g. the electron density, the column density, the covering factor, the ionisation parameter) should vary in such a way to keep the line properties constant.

\section{Conclusions}
\label{sec:conclusions}
In this paper we analysed the new \jwst/NIRSpec-IFU high- and low-spectral resolution observations of ULAS J1342, the second farthest QSO known so far. 
Our main findings are listed below:

\begin{itemize}
    
    \item By employing several emission lines, including the benchmark \Hb, and different SE scaling relations, we estimated the black hole mass for this source which ranges between $\rm \rm \log(M_{BH}/M_ {\odot})=8.6-9.3$ with a systematic uncertainty of the order of 0.4--0.5 dex. We also estimated \mbh\ using an alternative approach relying, for the first time with \jwst data, on an accretion disc modelling approach. This approach provided a black hole mass of $\rm \log(M_{BH}/M_{\odot})=9.2 \pm 0.2$ where the associated uncertainty is significantly smaller than that of single-epoch calibrations (0.2 dex against 0.4--0.5 dex). Here, we highlight that low-spectral resolution data of $z \gtrsim 4$ quasars allow for a clear description of the disc emission, therefore unlocking a new powerful way to investigate the masses of primeval quasars.

    \item By combining the black hole mass with the estimated bolometric luminosity, we derived a sub-Eddington accretion state ($\rm \lambda_{Edd}\sim 0.4$). Postulating an Eddington limited accretion, already over by the time of observations, the black hole seed should have been of the order of $\rm \sim 10^4$--$10^6 \, M_{\odot}$ at $\rm z\sim 15$--$25$. This mass range falls in the heavy seeds regime.
    
    \item The low-resolution data revealed the presence of extended \oiii, as well as two morphologically distinct \oiii emitters in the field of view, one of which in seeming interaction with the QSO galaxy. The extended \oiii halo reaches as far as $\sim$7 kpc from the nucleus. This structure is possibly tracing an outflow, a prospect suggested by the broad kinematics revealed by the high-resolution data, while another brighter \oiii outflow is detected in the nuclear region. The non detection of a high \oiii/[\ion{O}{ii}] ratio in the extended \oiii region argues against a massive presence of shocks in this region. The combined mass outflow rate spans the range 50--300 $\rm M_{\odot} \, yr^{-1}$ depending on the assumed density, and could be three times higher if the outflowing gas has a constant average volume density. These values overlap, in part, with those estimated for the star formation rate. Therefore the ionised mass expelled by the outflow could ultimately quench the star formation as early as during the epoch of reionisation. Additionally, a significant amount of mass and energy could be carried by other phases (e.g. fast nuclear, neutral, molecular) so that the total outflowing gas mass could easily surpass that consumed by star formation. 
    
    \item We found that several line ratios involving both optical and UV \Feii (\FeMg and \FeHb, measured here for the first time at $z>7$), widely employed to trace the BLR metallicity, are consistent with the average values observed at lower redshifts despite being produced at remote cosmic times. This, hints at an early chemical enrichment of BLRs in the first generation of quasars.

\end{itemize}

The analysis presented here on ULAS J1342 demonstrated, once again, the tremendous amount of information accessible only through \jwst. Despite being one of the most well-studied QSOs at high redshift, the new observations presented here revealed key features so far inaccessible. In particular, the synergy between the high- and low-spectral resolution observations proved key, both to detect the extended ionised gas halo, to perform the AD modelling and to robustly determine the gas kinematics. This work demonstrates how the new capabilities of \jwst allow self-consistently linking the nuclear and the galactic scales even as early as the epoch of reionisation.

\begin{acknowledgements}
We acknowledge M. Onoue for kindly providing the Gemini/GNIRS spectrum of ULAS J1342. This work is based on observations made with the NASA/ESA/CSA James Webb Space Telescope. The data were obtained from the Mikulski Archive for Space Telescopes at the Space Telescope Science Institute, which is operated by the Association of Universities for Research in Astronomy, Inc., under NASA contract NAS 5-03127 for JWST. These observations are associated with program 1219. BT, SC, GV and SZ acknowledge support by European Union’s HE ERC Starting Grant No. 101040227 - WINGS. AJB acknowledges funding from the "FirstGalaxies" Advanced Grant from the European Research Council (ERC) under the European Union’s Horizon 2020 research and innovation programme (Grant agreement No. 789056). FDE acknowledges support by the Science and Technology Facilities Council (STFC), by the ERC through Advanced Grant 695671 ``QUENCH'', and by the UKRI Frontier Research grant RISEandFALL. RM acknowledges support by the Science and Technology Facilities Council (STFC), by the ERC through Advanced Grant 695671 ``QUENCH'', and by the UKRI Frontier Research grant RISEandFALL. RM also acknowledges funding from a research professorship from the Royal Society. H\"U acknowledges funding by the European Union (ERC APEX, 101164796). Views and opinions expressed are however those of the authors only and do not necessarily reflect those of the European Union or the European Research Council Executive Agency. Neither the European Union nor the granting authority can be held responsible for them. 
SA, MP, and BRP acknowledge support from grants PID2021-127718NB-I00, RYC2023-044853-I, and PID2024-159902NA-I00, funded by the Spanish Ministry of Science and Innovation/State Agency of Research (MCIN/AEI/10.13039/501100011033) and El Fondo Social Europeo Plus FSE+. 
GC and EB acknowledge the support of the INAF Large Grant 2022 "The metal circle: a new sharp view of the baryon cycle up to Cosmic Dawn with the latest generation IFU facilities", the INAF GO grant ``A JWST/MIRI MIRACLE: Mid-IR Activity of Circumnuclear Line Emission''. EB acknowledges funding through the INAF ``Ricerca Fondamentale 2024'' program (mini-grant 1.05.24.07.01). IL acknowledges support from PRIN-MUR project “PROMETEUS”  financed by the European Union -  Next Generation EU, Mission 4 Component 1 CUP B53D23004750006. 

\end{acknowledgements}

\bibliographystyle{aa} 
\bibliography{bibl}

\begin{appendix}

\twocolumn

\section{Previous studies}
\label{app:previous_studies}

The multi-band properties of ULAS J1342 as well as its close environment have been the subject of several investigations. Ground based infrared spectroscopy of ULAS J1342 targeted the rest-frame UV wavelengths, allowing the detection of broad lines (e.g. \ion{Mg}{ii}$\lambda$2798, \ion{C}{iv}$\lambda$1549) suitable for estimating the black hole mass. In particular, the black hole mass based on \ion{Mg}{ii} SE calibrations is around $10^9$ \Msun with slight differences depending on the calibration employed and on the UV \ion{Fe}{ii} template adopted in the fitting procedure (\citealt{banados2018, onoue2020no}). Interestingly, the rest-frame UV spectral properties (continuum slope, line ratios) of ULAS J1342 align with typical values derived at lower redshifts (\citealt{onoue2020no}).

ULAS J1342+0928 is also one of the few X-ray detected sources at $z>7$. It was first observed with \chandra in a 45.1 ks campaign in 2017 (\citealt{banados2018chandra}), and again for 208 ks with \xmm in 2021 as part of the HYPERION XMM-Newton Heritage program (\citealt{zappacosta2023hyperluminous,tortosa2024hyperion}). These different observations measured consistent hard X-ray (between rest-frame 2--10 keV) luminosities, in particular $\rm 13.0^{+4.0}_{-3.4} \times 10^{44} \, erg \, s^{-1}$ in \citealt{banados2018chandra} against $\rm 17.1^{+3.1}_{-3.0}  \times 10^{44} \, erg \, s^{-1}$ in \citealt{zappacosta2023hyperluminous}. Yet, the photon index ($\Gamma$) was found to be steeper (at $\sim1.4 \, \sigma$) in the latter work ($\rm \Gamma = 2.87^{+0.43}_{-0.37}$ against $\rm \Gamma = 1.95^{+0.55}_{-0.53}$).

Taking advantage of NOEMA observations and employing the [\ion{C}{ii}]158$\mu$m-SFR scaling relations, \citet{venemans2017copious} classified ULAS J1342 as a highly star-forming galaxy, with a SFR in the range 85--545 \Msun $\rm yr^{-1}$ and a dynamical mass $<1.5 \times 10^{11}$ \Msun. Similar results concerning the host galaxy SFR were obtained using ALMA observations of the dust continuum ($\rm SFR = 150\pm30 \, M_{\odot}$; \citealt{novak2019alma}).
Further ALMA observations, centred on the [\ion{C}{ii}]158$\mu$m emission line (\citealt{banados2019z}), revealed the presence of two morphologically separate peaks and chaotic motions of the cold gas on kpc scales around the central quasar, suggesting an on-going galaxy merger whose gas reservoir is feeding the central quasar. Before the observational campaign described here, \jwst/NIRSpec observed this source in the fixed slit mode in both the G140H/F070LP and G235H/F170LP configurations. These observations were employed to study the absorbers along the line of sight to the quasar and investigate the evolution of the metal enrichment up to the epoch of reionisation in \citet{christensen2023metal}.

\section{Prism \Hb--\oiii region fit }
Here we show the spectral fit of the integrated prism spectrum extracted from a 6$\times$6 spaxels region centreed at the QSO location in the \Hb--\oiii region. The best fit model for the broad components and the continuum was employed for the PSF subtraction on a spaxel-by-spaxel basis.

\begin{figure}[h!]
\centering
\includegraphics[width=\linewidth,clip]{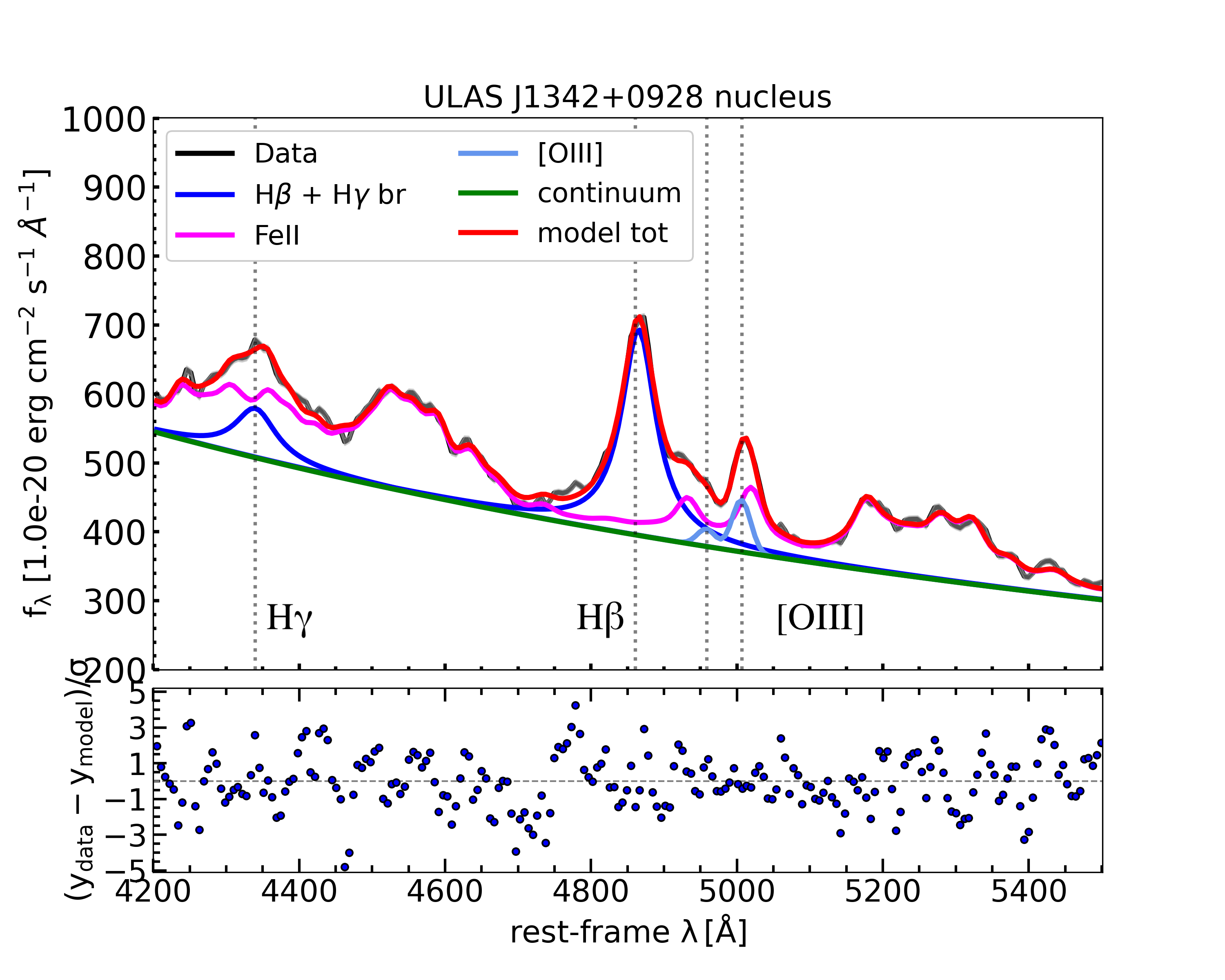}
\caption{Spectral fit of the \Hb--\oiii region of the low resolution data. The spectrum was extracted from the same region as the one shown in Fig.~\ref{fig:spec_fits_BLR}. All of the model components are colour-coded as in the legend.}

\label{fig:hb_fit_prism}
\end{figure}

\section{High resolution spectra of the \oiii emitters}
\label{app:o3emitters_spectra}
Here we show the high resolution spectra of the \oiii emitters detected in the field of view of ULAS J1342. In addition to the already shown spectra of the nuclear and extended \oiii halo, here we display those of the regions labelled as `NW', and `N' in Fig.~\ref{fig:o3_map_spectra}.

\begin{figure}[h!]
\centering
\includegraphics[width=\linewidth,clip]{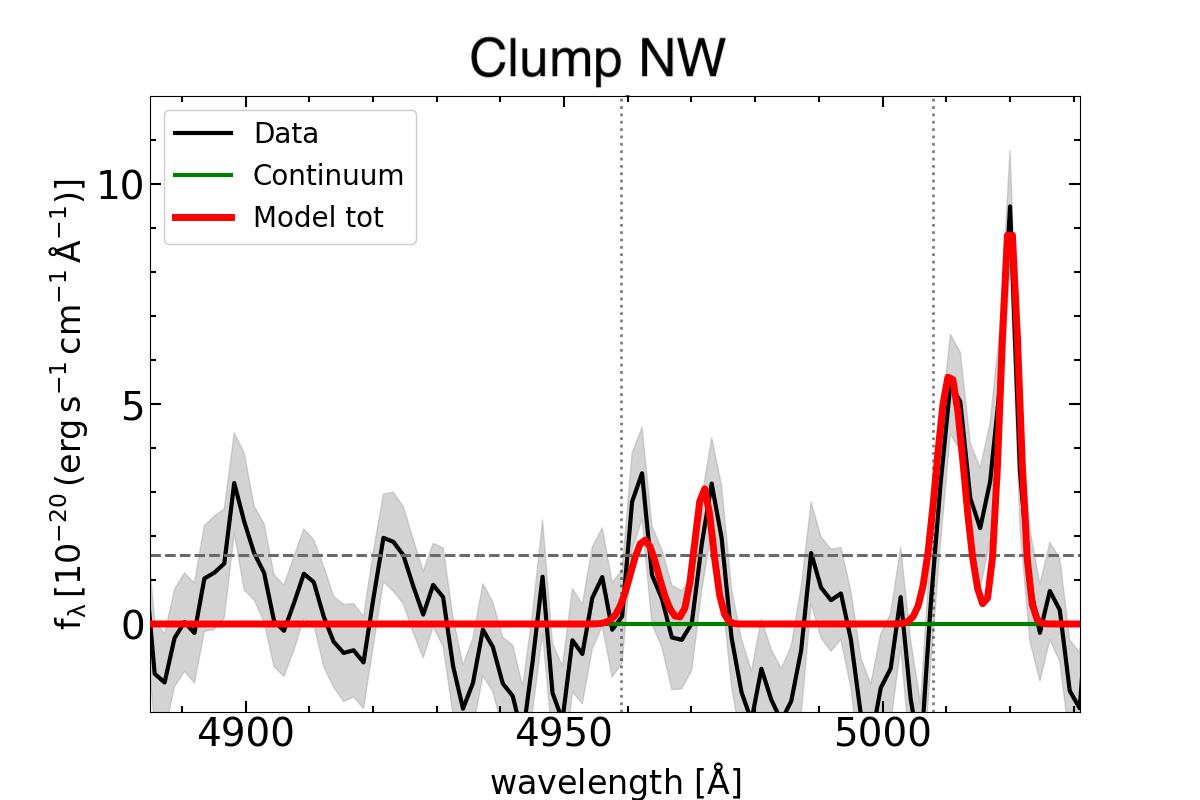}
\caption{Spectral fit of the high resolution spectrum extracted from the `NW' \oiii clump. The dotted lines mark the expected wavelengths of the \oiii lines at the redshift of the QSO.} 

\label{fig:NW_clump}
\end{figure}
\begin{figure}[h!]
\centering
\includegraphics[width=\linewidth,clip]{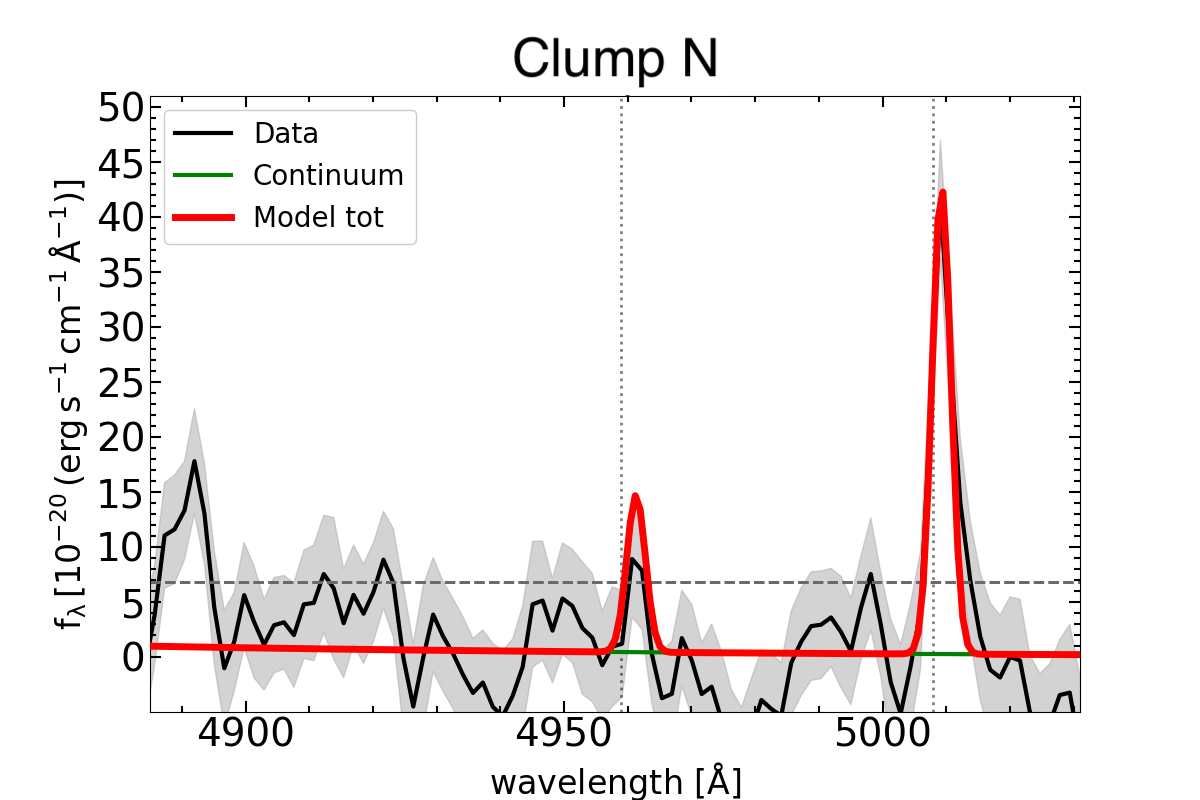}
\caption{Spectral fit of the high resolution spectrum extracted from the `N' \oiii clump. The dotted lines mark the expected wavelengths of the \oiii lines at the redshift of the QSO.} 

\label{fig:N_clump}
\end{figure}

\section{[\ion{O}{ii}]$\lambda$3728 map}
\label{app:o2map}
In this section, we show the flux map obtained by collapsing the low-resolution data-cube at the wavelengths covered by the [\ion{O}{ii}]$\lambda\lambda$3726,3729 ([\ion{O}{ii}]$\lambda$3728 for simplicity). The spatial distribution of this line is shown as dashed contours. There, it can be seen that, while there is a general co-spatiality between the \oiii and the [\ion{O}{ii}], in the nucleus, and in the other \oiii emitters, this is not the case for the extended SW region. In principle, the ionisation parameter in the \oiii halo can be different from that of the nuclear region, therefore implying a different \oiii/[\ion{O}{ii}] ratio. If we suppose that the ratio between \oiii and [\ion{O}{ii}] (O32) is the same as in the nuclear region, the most likely cause for the non detection of [\ion{O}{ii}] along the \oiii plume is the sensitivity. Assuming the O32 ratio to be the same as in the nuclear region ($\sim$3.1) the expected [\ion{O}{ii}] flux, given the measured \oiii in the extended region, would fall below the 2$\sigma$ contours reported here. We can use the measured \oiii flux and the [\ion{O}{ii}] RMS to set a lower limit for the O32 ratio which is O32$\rm _{min}$ $\sim$1. This ratio helps in constraining the properties of the extended \oiii emission, as it does not seem consistent with shock-induced photoionisation. Assuming, for instance the shock line ratios calculated by \citet{allen2008mappings}, we would expect, for any shock velocity $\lesssim$1,000 km s$^{-1}$, to have O32$\gtrsim$1. Therefore shocks do not appear as a main agent in the photoionisation in this region.

\begin{figure}[h!]
\centering
\includegraphics[width=\linewidth,clip]{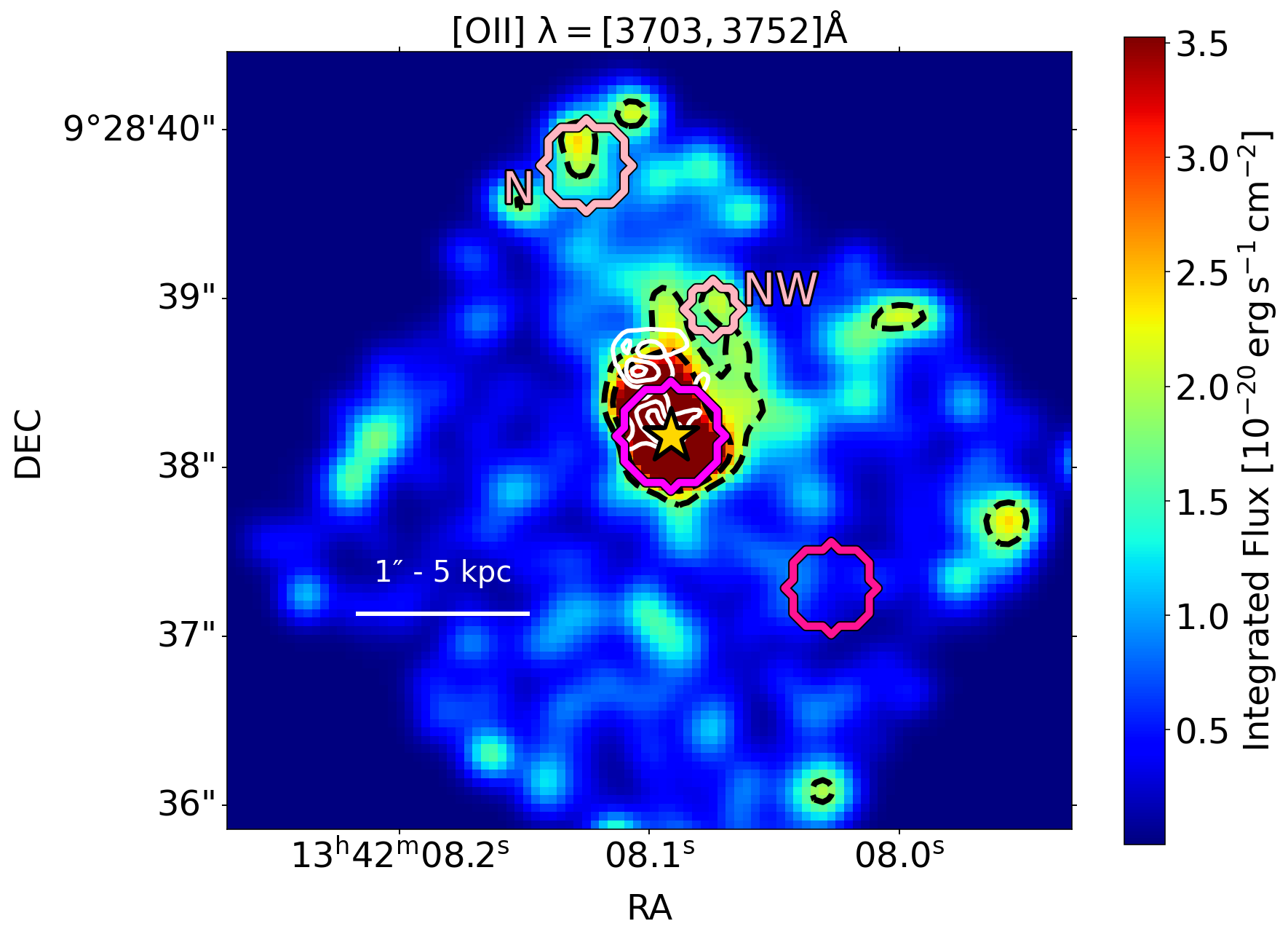}
\caption{Flux map of the [\ion{O}{ii}]$\lambda 3728$ map. Dashed lines mark respectively 2 and 3 $\sigma$ contours. White contours represent the same ALMA contours described in Fig. \ref{fig:o3_map_spectra}. The coloured regions are the same as in Fig. \ref{fig:o3_map_spectra}.}

\label{fig:o2_map}
\end{figure}

\section{Possible systematics on the broad line ratios}
\label{app:fe2mg2_systematics}
Here we briefly explore the possible systematics affecting the comparison between the \FeHb and the \FeMg ratios shown in Fig. \ref{fig:RFe_z} among different authors. The choice of the \ion{Fe}{ii} templates plays a key role to estimate the aforementioned ratios. Indeed, the emission of the individual Fe multiplets, which is fixed in the case of empirical templates (e.g. \citealt{vestergaard2001empirical, tsuzuki2006fe}) constrains the width and the flux of the \ion{Mg}{ii} and \Hb lines (e.g. \citealt{shin2019fe, wang2022metallicity, lai2024xqz5}), therefore crucially affecting the final ratios. A more quantitative exploration of this topic is given in Sect. 5.3 of \citet{shin2019fe}. Additionally, also the choice of the local continuum as well as the line profile chosen to model the \Hb and the \ion{Mg}{ii} play an important role. Different choices of these parameters among different works could introduce systematic differences. Since the average \FeHb and \FeMg ratios do not appear to clearly evolve with the redshift, it is, however, hard to speculate that these possible systematics combine in such a way to exactly counteract an intrinsic decrease in these broad line ratios. 

Concerning ULAS J1342, we also highlight that analogous measurements of the \ion{Fe}{ii}\textsubscript{UV}/\ion{Mg}{ii} ratio in this source had already been performed in \citet{onoue2020no} and \citet{yang2021probing}. The difference in the measurements shown in the top panel of Fig.~\ref{fig:RFe_z} encapsulates the contributions of the different assumptions on the \ion{Mg}{ii} line profile (Gaussian, Lorentzian), the different \ion{Fe}{ii} templates employed in the fit, as well as the possible (yet minor) effect of variability.

\section{The effect of extinction}
\label{app:extinction}

The presence of gas and dust along the line of sight within the BLR and/or the host galaxy could bias our estimates of black hole mass and bolometric luminosity. In the case of the AD modelling, assuming a non-greybody like extinction curve, reddening would make the observed SED dimmer and redder (i.e. colder) than the actual one. This would translate into a lower disc luminosity and a higher black hole mass than the intrinsic. Notwithstanding the points presented in the main text argue that reddening is not a major issue for our source, we attempted to estimate how the possible presence of reddening would alter our \mbh\ and \lbol\ estimates. To this aim, we directly introduced intrinsic reddening in our AD modelling.

The magnitude of the extinction might be degenerate with both \mbh\ and \lbol. As we wish to avoid unphysically high values of E(B-V), inconsistent with the previous considerations, we need to inform our fitting routine about the E(B-V) distribution actually observed in blue QSOs. We achieved this by taking advantage of the QSO compilation assembled in \citet{krawczyk2015mining}, who explored the distribution of reddening in $\sim$35,000 sources between $0<z<5.3$. In brief, we selected all the sources in their sample between $\pm 0.25$ dex from the $\rm L_{3,000\AA}$ measured in ULAS J1342 ($\sim 1,630$ sources) and computed the E(B-V) distribution. We then fitted the E(B-V) distribution, only including positive values\footnote{Negative E(B-V), would imply sources with a bluer continuum than the one assumed to be unextinguished. Albeit such sources are interesting to understand the shape of the unobscured continuum, including those sources would make the average E(B-V) slightly smaller. Because of this, our approach to avoid such objects gives more conservative constrains on the fitted parameters.} using a half Gaussian function. We used these distributions as priors for the E(B-V) values entering in the model likelihood. A Small Magellanic Cloud (SMC) extinction curve has been found to well reproduce the outliers in the QSO colour distribution (e.g. \citealt{hopkins2004dust, krawczyk2015mining}), therefore we adhered to such prescription. We used the `G24\_SMCAvg' extinction curve from the Python package `dust\_extinction' (\citealt{gordon2024dust_extinction}) with a total-to-selective ratio of $\rm R_V=3.0$ (\citealt{gordon2024expanded}) to redden the intrinsic SED and proceeded to perform the AD fits in the same way as described in the main text. The accretion parameters slightly shift to $\rm \log (M_{BH}/M_{\odot})=9.0\pm0.4$ and $\rm \log(L_{bol} / L_{\odot})=47.0\pm0.3$. As a result of the mild decrease in \mbh\ and the increase in \lbol, $\rm \lambda_{Edd}$ increases to 0.8, yet ULAS J1342 remains below the Eddington limit.

\end{appendix}

\end{document}